# Constraints in Random Effects Age-Period-Cohort Models*

Liying Luo
*Department of Sociology and Criminology*
*Population Research Center*
*Pennsylvania State University*

James S. Hodges
*Division of Biostatistics, School of Public Health*
*University of Minnesota*

March 2019

*An earlier version of this article was presented at the annual meetings of the Population Association of America (April 2017, Chicago). The project benefited from support provided by the Population Research Institute. We thank Wayne Osgood for his helpful comments. Please direct correspondence to Liying Luo, 202 Oswald Tower, University Park, PA 16802 or email liyingluo@psu.edu.

**Abstract**

Random effects (RE) models have been widely used to study the contextual effects of structures such as neighborhood or school. The RE approach has recently been applied to age-period-cohort (APC) models that are unidentified because the predictors are exactly linearly dependent. However, it has not been fully understood how the RE specification identifies these otherwise unidentified APC models. We address this challenge by first making explicit that RE-APC models have greater—not less—rank deficiency than the traditional fixed-effects model, followed by two empirical examples. We then provide intuition and a mathematical proof to explain that for APC models with one RE, treating one effect as an RE is equivalent to constraining the estimates of that effect's linear component and the random intercept to be zero. For APC models with two RE's, the effective constraints implied by the model depend on the true (i.e., in the data-generating mechanism) non-linear components of the effects that are modeled as RE's, so that the estimated *linear* components of the RE's are determined by the true *non-linear* components of those effects. In conclusion, RE-APC models impose arbitrary though highly obscure constraints and thus do not differ qualitatively from other constrained APC estimators.

# 1. Introduction

In a statistical model, the effects of a predictor can be modeled as either random or fixed, depending on the researcher's goal, the study design, and the data structure. A random effect (RE) model[1] is useful when it captures otherwise unexplained variation associated with context and environment and thus permits estimates of both group-level and individual-level characteristics to be more accurate and efficient (Gelman and Hill 2007; Raudenbush and Bryk 2002).

The RE approach has recently been applied to age-period-cohort (APC) models (see, e.g., Yang and Land 2006, 2008), in which researchers attempt to estimate the association between the outcome and a person's age (A), the time period (P), and the person's birth cohort (C), respectively. It is well-known that because the three predictors A, P, and C are *exactly* linearly dependent, APC models are not identified, meaning that no unique set of estimates fits the data best (Fienberg and Mason 1979; Mason and Fienberg 1985; Rodgers 1982). The consensus among methodologists is that one should rely on theory to choose one solution from the infinitely many solutions to the estimating equation (Fienberg 2013; Luo 2013b, 2013a; Luo et al. 2016; O'Brien 2013, 2017).

Contrary to this conventional wisdom, some have noted that modeling one or more of the predictors as an RE identifies the otherwise unidentified APC model without appearing to require the researcher to impose a constraint. For example, a series of articles has stated that when modeling the P and C effects as random effects, APC models become identifiable, that is, a researcher obtains uniquely-defined estimates for A, P, and

---

[1] Certain random effect models are also called mixed effects models, multilevel models, or hierarchical models.

C effects that reveal the underlying aging, social change, and cohort process. This method has subsequently been applied to a variety of topics including mortality (Masters, Hummer, and Powers 2012), happiness (Yang 2008), social trust (Gauchat 2012; Robinson and Jackson 2001; Schwadel 2010), gender attitudes (Shu and Zhu 2012), voting (Smets and Neundorf 2014), and environmental spending (Pampel and Hunter 2012).

Methodologists have begun to raise questions about the RE approach to the APC problem. For example, Bell and Jones (2018) showed that the data structure (e.g., the number of A, P, and C categories included in the data) affects the estimates from the RE-APC models markedly. For another example, O'Brien (2017) conjectured that assuming the effect of A, P, or C to be random produces a result similar to constraining that effect's linear trend to be near zero.

However, besides the above observations, more methodological studies are needed about how exactly the RE specification identifies APC models. The present paper contributes to this literature in three important ways. First, we clarify that APC models with one RE (hereafter 1-RE-APC models) in fact impose multiple constraints, not just one constraint as in traditional FE-APC models, but one on the linear component and another on the intercept (or level) estimate of the effect that is considered random in the model.

Second, following an empirical demonstration and some intuition, we provide a mathematical proof that the 1-RE-APC model constrains the estimated linear component of the random effect to be zero. This sharpens the observation made by O'Brien (2017) and noted above. Although previous research has noted the problems in 1-RE-APC models, there has been no formal exposition about why they necessarily impose such con-straints. We address this gap in the literature by providing a much-needed elaboration. We also provide exemplary R code in Appendix A so that interested readers can see the results for themselves.

Third, we clarify misunderstandings about the constraints in APC models with two RE's (hereafter 2-RE-APC models). 2-RE-APC models may appear to impose constraints on the two RE's similar to the constraint that 1-RE-APC models imposes on one effect, but we show that in fact a 2-RE-APC model imposes a different type of constraint that depends on the size of the non-linear components of the true effects that have been modeled as random. This constraint is in addition to Bell and Jones's (2018) observation about the model's dependence on the number of levels that each effect has. Although such constraints do not have a simple closed-form expression, we document the dependence of a 2-RE-APC model's constraints on the true non-linear components of the effects modeled as random and explain how this dependence works.

The paper is organized as follows. We begin by comparing a traditional FE-APC model and an RE-APC model to show the additional rank deficiency in the latter. Next, we provide empirical examples in which the RE-APC model results are highly sensitive to the choice of effects that are modeled as random. We then provide an intuitive explanation and a mathematical proof of how and why 1-RE-APC models constrain the random effect's estimated linear and intercept components to be zero. For 2-RE-APC models, we document and explain how the estimated A, P, and C effects depend on the size of the non-linear components of the effects modeled as random. We conclude that the RE-APC specification does in fact impose constraints, albeit obscure ones, and thus is qualitatively no different from other constrained estimators.

## 2. FE- and RE-APC Models: Rank Deficiency and Estimation Procedure

To provide context and define notation for a later explanation about how an RE-APC model adds constraints implicitly, we begin with the traditional FE-only APC model, show how the design matrix changes when some effects are modeled as random, and then introduce the machinery that produces estimates for an RE model.

### 2.1 The Traditional FE-APC Model

In traditional APC analyses, the A, P, and C effects are all modeled as FE's. The standard matrix form of an FE-APC model is:

$$y = X\beta + \varepsilon, \qquad (1)$$

where $y$ is a vector of outcomes; $X$ is the design matrix for the FE's; $\beta$ is the vector of FE coefficients; and $\varepsilon$ denotes normally distributed random errors with mean 0 and variance $\sigma_e^2$. To illustrate what $X$ looks like, suppose that one models a dataset with three categorical predictors A, P, and C and suppose that A and P have three groups each so that C has five groups, denoted as $a = 3$, $p = 3$, and $c = 5$, respectively. For each A-P combination, there is one observation, so the total number of observations in the dataset is nine. Table 1 gives the design matrix $X$ for an FE-only APC model. If more A, P, and C categories or more than one observation in each A-P combination are included, Table 1 can be expanded, with a new category or new observations implying additional columns or rows respectively in the matrix $X$. The predictors can be treated as if they are continuous predictors, which does not affect the discussion subsequently.

[Table 1 about here]

With the last group omitted for A, P, and C each, the FE design matrix $X$ consists of nine columns. However, $X$ has only eight linearly independent columns, so it is one less than full rank. As a result, **X'X** is not invertible and the FE-APC model is unidentified, that is, no uniquely-defined solution exists because an infinite number of solutions give *identical* fits to the data.

The identification problem can be more accurately characterized by breaking the effects of A, P, and C into linear and non-linear (curvature or deviations from linearity) components. It has been known at least since Holford (1983) that the linear components of the A, P, and C effects are exactly dependent and thus cannot be estimated without constraints while non-linear A, P, and C effects are identified and thus can be estimated without constraints.

Specifically, the linear dependency among A, P, and C implies that there is a non-zero vector $b_0$ such that the product of the design matrix and $b_0$ equals zero:

$$\boldsymbol{X}b_0 = 0. \tag{2}$$

In other words, $b_0$ represents the null space of the design matrix **X**, which has dimension equal to one. Kupper and colleagues (1985) provided a closed-form representation for $b_0$, making it clear why the linear components of A, P, and C effects are unidentified.

**2.2 The RE Approach**

We now begin to address the problems with RE-APC models by showing how the design matrix changes when one or two of the effects are modeled as random. To understand the implication of an RE specification for APC models, we begin by describing the RE model in general. The standard matrix form of a simple RE model is

$$y = \boldsymbol{W}\beta + \boldsymbol{Z}u + \varepsilon, \tag{3}$$

where $y$ is a vector of outcomes; **W** and **Z** are the design matrices for the FE's and RE's, respectively; $\beta$ is the vector of FE regression coefficients; $u = (u_1, u_2, \ldots u_j, \ldots, u_J)$ de-notes the vector of RE's with each $u_j$ modeled as normally distributed around a zero mean, i.e., $u \sim i.i.d. N(0, \sigma_u^2)$; and $\varepsilon$ denotes normally distributed random errors with mean 0 and variance $\sigma_e^2$.

To specify an RE-APC model, the researcher needs to decide (1) how many effects (one or two) and (2) which effects will be modeled as RE's. Then the corresponding FE and RE design matrices, **W** and **Z** respectively, can be specified. In the discussion below, without loss of generality we focus on two RE-APC models: a 1-RE-APC model that treats the C effect as the sole RE and a 2-RE-APC model that treats P and C effects as RE's.

Consider the same example as in Section 2.1, with three A groups, three P intervals, and five C's. Table 2 gives the design matrix for a 1-RE-APC model that treats the A and P effects as FE's and the C effect as the sole RE. The combined design matrix (**W** | **Z**), where the symbol "|" indicates concatenation of the matrices **W** and **Z** to create a single matrix, has 10 columns, while the FE design matrix **X** in Table 1 has nine columns. That is, (**W** | **Z**) has one more column than **X** in the columns for the C effects. Like the FE design matrix **X**, because of the linear dependence, there are only eight independent columns in (**W** | **Z**). That is, the design matrix (**W** | **Z**) for 1-RE-APC models is *two* less than full rank, in contrast to the FE design matrix **X**, which has rank one less than full.

[Table 2 about here]

Table 3 gives the design matrix for a 2-RE-APC model that treats the A effect as an FE and the P and C effects as RE's. The combined design matrix (**W** | **Z**) in Table 3 now has 11 columns, one more than for the 1-RE-APC model above and two more than

for the FE-APC model. This combined design matrix (**W** | **Z**) is *three* less than full rank because there are still only eight independent columns.

[Table 3 about here]

Why do RE-APC models suffer from *greater* rank deficiency than the FE-APC model? The redundancy added by making an effect an RE can be understood as follows: for each RE, add together its design-matrix columns; the result is identical to the design column for the FE intercept. For example, adding together Table 2's columns corresponding to the C effects $\gamma_1$ through $\gamma_5$ gives a vector of 1's, which is identical to the column corresponding to the FE intercept, Table 2's first column. Similarly, in Table 3 the result from adding together the columns corresponding to $\beta_1$ through $\beta_5$ or corresponding to $\gamma_1$ through $\gamma_5$ is also identical to the FE intercept column. That is, modeling an effect as random adds a redundant intercept for that effect without affecting the fact that the linear components of the effects are not identified. This implies that modeling an effect as random creates an even greater need for constraints than in the FE-only APC model, which requires just one constraint (e.g., constraining two age groups to be equal) in addition to the usual constraints that define the FE parameterization (i.e., the sum-to-zero constraint or declaring one group to be the reference group).

**2.3 Estimating FE- and RE-APC Models**

Although as we illustrated above neither the FE- and RE-APC models have a full-rank design matrix, their rank deficiencies are handled differently in conventional statistical software such as R, SAS, and STATA. For an FE-APC model fitted as if it were a classical regression model, the software either gives an error message and does not provide results or it arbitrarily drops one design matrix column and estimates the resulting model. This occurs because when the model is estimated by minimizing the residual sum of squares:

$$(y - X\beta)'(y - X\beta) \tag{4}$$

the minimizing $\beta$ is given by

$$[X'X]^{-1}X'y, \tag{5}$$

and the matrix $X'X$ does not have an inverse due to the linear dependence.

By contrast, when at least one effect is modeled as random, in the conventional (i.e., non-Bayesian) analysis of this model for outcomes treated as normally distributed, statistical software provides estimates and inference results for all three of A, P, and C without complaints. To understand this difference from a FE-APC fit, one needs to know how RE models are estimated. Estimating an RE model in such software has two steps, which we briefly describe below; Sections 4 and 5 will discuss in more detail how the RE machinery uses the data, as it pertains to RE-APC models.

Step 1: The unknown variances (the error variance and all RE variances) are estimated by maximizing the restricted likelihood (RL, also known as the residual likelihood).

Step 2: The estimates of the unknown variances are now taken as if known to be true and $\beta$ and $u$ are estimated by maximizing the penalized likelihood, which for normally distributed outcomes is identical to choosing $\beta$ and $u$ to minimize the following equation:

$$(y - W\beta - Zu)'(y - W\beta - Zu) + \hat{\sigma}_e^2 u'\widehat{G}^{-1}u, \tag{6}$$

where $\hat{\sigma}_e^2$ is the estimated error variance and $\widehat{G}$ the estimated RE covariance matrix. Solving Equation (6) has the effect of replacing the matrix $X'X$ in Equation (5) with:

$$\begin{pmatrix} W'W & W'Z \\ Z'W & Z'Z + \hat{\sigma}_e^2 \hat{G}^{-1} \end{pmatrix} \tag{7}$$

and replacing $X'y$ with $[W|Z]'y$.

For the RE models considered in this paper, the bottom right block in Matrix (7), $Z'Z + \hat{\sigma}_e^2 \hat{G}^{-1}$, is invertible so that the estimate of $(\beta, u)$ is uniquely determined, when taking as known the estimates of the error variance $\sigma_e^2$ and any unknowns in $G$. The second term in (6), $\hat{\sigma}_e^2 u' \hat{G}^{-1} u$, is sometimes called a penalty, as in "penalized likelihood" or "penalized regression". For identified models, this term has the effect of increasing the residual sum of squares $(y - W\beta - Zu)'(y - W\beta - Zu)$ when $u$ is not zero and thus constraining the estimates of the RE's, including linear and non-linear effects, by shrinking them toward zero (i.e., the center of $u$'s distribution). As the statistical literature on RE models has noted, although introducing the penalty $\hat{\sigma}_e^2 u' \hat{G}^{-1} u$ creates a bias in the estimate of $(\beta, u)$, that may be desirable on balance because it can reduce the variance of the estimated $u$ by enough to reduce the mean squared error of the estimate of $(\beta, u)$ (which is the sum bias$^2$ + variance; see, e.g., Gelman and Hill 2007).

The above description of the two-step estimation procedure offers important clues about why statistical packages can provide estimates and inference results for the A, P, and C effects without complaint in an RE specification. Some scholars mistakenly take it to mean that the RE approach "overcomes" or "alleviates" the identification problem (see, e.g., Gauchat 2012; Jaacks et al. 2013; Schwadel 2010) without requiring researchers to make subjective assumptions. However, as we will demonstrate and explain in the rest of the paper, this optimistic view is unwarranted. Obtaining a set of estimates does not automatically mean that the identification problem has been solved without additional assumptions or that the resulting estimates are meaningful, useful, or reliable. In the next section, we demonstrate the identification problem in RE-APC models and the implications of the RE specification for effects estimates using two empirical examples. We offer formal proofs and explanations in Sections 4 and 5.

## 3. Empirical Examples: Implications of the RE Assumption for APC Models

Although later sections give mathematical proofs and computational results about the constraints in RE-APC models, such evidence may be highly technical and abstract even for *Sociological Methodology* readers. Therefore, we begin with two empirical examples, analyzing individual-level data from the General Social Survey (GSS) data to make a less technical argument. We provide theoretical exposition in Sections 4 and 5.

Since the early 1970s, the GSS has collected data on a variety of topics including attitudes, behaviors, and attributes; scholars have frequently used GSS data to document and explain changes in American society. We chose two GSS measures, self-reported happiness level and conservative-liberal views, for which temporal trends have been a frequent subject of social science research and APC analysis (see, e.g., Davis 2004, 2013; Glass 1992; Hagerty and Veenhoven 2003; Hout and Fischer 2014; Rodgers 1982; Smith 1990; Yang 2008). Specifically, for every year or every other year since 1974, the GSS has asked participants to report their happiness on a three-level scale: happy, pretty happy, and not too happy. For simplicity and ease of interpretation, we grouped the categories "happy" and "pretty happy", which results in a binary outcome "happy" versus "not too happy". To measure political ideology, the GSS asked participants to describe their political views on a seven-point scale with one being extremely liberal and seven extremely conservative.

For both examples, we selected GSS participants interviewed between 1974 and 2014 and aged between 18 to 74 to minimize the number of age–period combinations with too few observations for the elders. The predictors in all of the analyses that follow are the participant's age, year of survey, and cohort membership. Because our goal is not to provide a full understanding of the social determinants of happiness or conservative-liberal views but merely to illustrate some methodological points about the RE assumption in unidentified models, we do not include typical covariates such as gender, education, and ethnicity. We omitted individuals with missing information on the outcome or predictors, giving sample sizes of 54,674 and 46,591 for the happiness and conservative-liberal views analyses, respectively. We analyzed these samples from the 1974-2014 GSS data using six RE-APC models, each having one or two of the A, P, or C effects modeled as random. We used five-year interval width for all A, P, and C groups to avoid additional complications of multiple block constraints arising from unequal interval widths for these groups (Luo and Hodges 2016). Tables 1A and 2A present the estimation results for the six RE-APC models.

Figures 1 and 2 illustrate the immediate implications of an RE specification for effect estimates for the happiness and political views examples, respectively. For both examples, not only do the A, P, and C effects estimates differ between these models, their overall *patterns* differ markedly between models depending on whether an effect is modeled as random or fixed. For example, the three RE models with a random C effect suggest that the odds of being happy increased with age up to age 30, remained at the same level until about 55, and began to rise again in older ages; and that Americans were less happy in the most recent decade after few changes for about 15 years, with little differ-ence between older and younger cohorts. However, in the other three RE models with a fixed C effect, the age effects on happiness show an inverted U shape with a plateau between the late 20s and early 50s; happiness did not change over the years between 1974 and 2014; and younger cohorts were less happy than older cohorts.

[Figures 1 and 2 about here]

Similarly, the estimates of A, P, and C effects on political views differ across the six models so that even the general patterns in these effect estimates depend on which effects are modeled as RE's. For example, the model with random A effects and fixed P and C effects (Model 1) suggests a substantial growth in conservative political views from 1974 to 2014, with an inter-cohort decline and largely flat age pattern. In contrast, the RE-APC model with a fixed A effect and random P and C effects (Model 4) indicates that American people tended to be more conservative as they grew older and were more liberal in the 1980s, 1990s, and 2000s than in the 70s and 2010s; the baby boom cohorts, born between mid-1940s and mid-1950s, were more liberal than older and younger cohorts. That is, researchers interested in temporal trends in political ideology in the United States would reach different substantive conclusions about changes in political views depending on the essentially arbitrary choice of which effects to model as random effects.

The above empirical examples confirm O'Brien's (2017) observation based on aggregated data that the A, P, and C effects on tuberculosis rates change with the choice of which effect is considered random. Such empirical evidence and the exposition in Section 2 suggest that the RE approach must impose some type of constraints, albeit implicit, to identify the otherwise unidentified APC models. However, two questions are yet to be answered: exactly what constraints do the RE-APC models impose and how can such constraints be accurately described? In the following sections, we address these questions by first showing and proving the constraints implicit in APC models with one RE. Because the constraints in APC models with two RE's do not have a closed-form expression, we discuss this issue by documenting the behavior of estimates from such models under various circumstances.

**4. Constraints in APC Models with One Random Effect**

We first present computational results that illustrate the mathematical results and provide important clues about the constraints in 1-RE-APC models in which one of the three effects is treated as random. Such results offer direct evidence for the assertion that using a 1-RE-APC model is equivalent to constraining the estimates of both the linear component and the intercept of the random effect to be zero. We provide an intuitive explanation and a mathematical proof of this assertion in Section 4.2.

**4.1 Computational Results for 1-RE-APC Models**

We describe a computation method and results for the model with A and P modeled as FE's and C modeled as an RE. As will become clear, modeling either A or P as the sole RE does not differ in any material way.

We derive for any 1-RE-APC design the linear combination of the data that estimates the C effect's intercept and its linear component. Specifically, as in previous equations, $\mathbf{W}$ and $\mathbf{Z}$ are the design matrices for the FE's and RE's respectively, each with $a \cdot p$ rows and $a + p - 1$ columns. Define the combined design matrix as $\mathbf{Q} = (\mathbf{W} \mid \mathbf{Z})$. Then for a given ratio of the error and RE variances, $\lambda = \sigma_e^2 / \sigma_c^2$, the estimate of the vector of effects is

$$\hat{\theta} = (\mathbf{Q}'\mathbf{Q} + \lambda \mathbf{D})^{-1}\mathbf{Q}'y = \mathbf{M}y \text{ for } \mathbf{M} = (\mathbf{Q}'\mathbf{Q} + \lambda \mathbf{D})^{-1}\mathbf{Q}', \qquad (8)$$

where **D** is a diagonal matrix with $a + p - 1$ diagonal entries of 0 corresponding to the intercept, age FE's, and period FE's, followed by $a + p - 1$ diagonal entries of 1 corresponding to the cohort RE. The first $a + p - 1$ elements of $\hat{\theta}$ are thus the estimated intercept, A effects, and P effects; the last $a + p - 1$ elements are the estimated C effects.

The level (or intercept), linear component, quadratic component, and higher order components of the C effect are estimated using linear functions of $\hat{\theta}$. Specifically, the C effect's intercept is estimated as $\alpha'\hat{\theta} = \alpha'\mathbf{M}y$, where $\alpha' = (0, ..., 0, 1, ..., 1)/(a + p - 1)$, where $\alpha$ has $a + p - 1$ entries that are 0 followed by $a + p - 1$ entries that are 1. Similarly, the C effect's linear component is estimated as $\beta'\hat{\theta} = \beta'\mathbf{M}y$ where $\beta' = (0, ..., 0, \beta_1)$ and $\beta_1 = (b'b)^{-1}b'$ for $b' = (1, 2, ..., a + p - 1) - (a + p - 1)(a + p)/2$.

The logic of the computation is thus as follows: for any design, i.e., any specification of $a$ age groups and $p$ period groups, and any value of the error-RE variance ratio $\lambda$, to verify that the C effect's intercept and linear components are estimated to be zero irrespective of the data $y$, one can compute **M** and thus $\alpha'\mathbf{M}$ and $\beta'\mathbf{M}$ to examine whether $\alpha'\mathbf{M} = \beta'\mathbf{M} = 0$.

We thus computed $\alpha'\mathbf{M}$ and $\beta'\mathbf{M}$ for designs with the number of A and P groups, i.e., $a$ and $p$, between three and 30 inclusive and for seven values of $\lambda$, 0.001, 0.01, 0.1, 1, 10, 100, 1000, covering the range of shrinkage of the random C effect from almost no shrinkage to shrinkage nearly to zero respectively. For each combination of $a$, $p$, and $\lambda$, we set up the combined design matrix **Q**, compute **M**, compute $\alpha'\mathbf{M}$ and $\beta'\mathbf{M}$, and save the entries in $\alpha'\mathbf{M}$ and $\beta'\mathbf{M}$ that are largest in absolute value. Appendix A gives details and R code.

Table 4 shows the computational results obtained by running this code on a Macintosh laptop using R version 3.5.2, for each combination of $a$, $p$, and $\lambda$. The figures in the columns are respectively $a$, $p$, $\lambda$, the largest absolute value of entries in $\alpha'M$ (for estimating the intercept of the C effects), and the largest absolute value of entries in $\beta'M$ (for estimating the linear component of the C effects). All of the $\alpha'M$ are $2 \cdot 10^{-12}$ or smaller and all of the $\beta'M$ are $4 \cdot 10^{-13}$ or smaller, numbers that are numerically equivalent to zero[2]. Thus, for any APC model included in this computation, the linear component and intercept of the C effect are estimated to have magnitude at most very close to zero, regardless of the number of A and P groups and thus C groups, regardless of the magnitude of shrinkage, and regardless of the data $y$.

[Table 4 about here]

These computational results show that for any dataset not only the linear component of the random effect will be estimated to be effectively zero, as O'Brien (2017) suggested, but the intercept (or average level) will also be estimated to be zero. It seems extremely unlikely that a researcher would choose these constraints on the cohort effect, if they were understood and explicit. The following section provides intuition and a math-ematical proof that apart from tiny errors arising from doing algebra by computer, these estimates are exactly zero.

---

[2] Linear-algebra software does not do exact matrix inversions; it computes matrix inverses to within the machine's accuracy. Results such as those presented here are well known to be equivalent to zero. If one finds this unsatisfying, note that the estimated linear component of the cohort effect will be smaller in absolute value than $4 \cdot 10^{-13} \cdot (a + p - 1) \cdot y_{max}$ where $y_{max}$ is the observation ($y$) that is largest in absolute value. This is effectively zero relative to $y$. Running this code using other machines or other versions of R may give results that differ slightly but are still all numerically equivalent to zero.

**4.2 Explaining the Constraints in 1-RE-APC Models: Some Intuition and a Mathematical Proof**

Why does assuming an effect to be random in a 1-RE-APC models imply that the estimated linear component and intercept of the RE are necessarily zero? We now provide some intuition and a mathematical proof for this conclusion. Although the explanation and proof focus on the model in which C is the sole RE, an analogous proof for a model with A or P as the sole RE follows immediately.

Intuition for the results in Section 4.1 is as follows. Assume the random C effect is parameterized in the standard manner, with one level of the RE for each cohort. The intuition (as well as the proof to be provided later) allow any parameterization of the fixed A and P effects. $\mathbf{W}$, $\mathbf{Z}$, $\beta$, and $u$ are as defined in Equation (3), and $\lambda$ just above Equation (8). The A, P, and C effects are estimated by solving the following problem: Choose the A and P effects in $\beta$ and the C effect in $u$ to minimize the penalized residual sum of squares[3]:

$$L(\beta, u) = (y - \mathbf{W}\beta - \mathbf{Z}u)'(y - \mathbf{W}\beta - \mathbf{Z}u) + \lambda u'u, \qquad (9)$$

where $\lambda$ is the ratio of the error and RE variances, $\lambda = \sigma_e^2 / \sigma_c^2$, and take $\lambda > 0$ as given. We will show that for any $(\beta, u)$ in which either or both of the C effect's level (intercept) and linear component are non-zero, $L(\beta, u)$ can be made smaller by setting C's level and linear component to zero and adjusting $\beta$ so that the model fit and thus the residuals $(y - W\beta - Zu)$ do not change. The latter is possible because the C effect's level is redundant with the FE intercept in $\beta$ and the C effect's linear component is redundant with the linear components of A and P. Therefore, the C effect's estimated level and linear component must be zero; otherwise, $L(\beta, u)$ could be made smaller by setting them to zero and adjusting $\beta$ accordingly.

---

[3] The penalized residual sum of squares is equivalent to the penalized likelihood for a model with normal errors and normally-distributed random effects.

Now we provide a formal proof of the above assertion. The steps in the proof are as follows. First, reparameterize the A and P effects from $\beta$ to $\beta^*$, in which the elements of the new parameter $\beta^*$ are the linear, quadratic, etc. components of the A and P effects. Second, reparameterize the C effect from $u$ to $u^*$, in which the elements of the new parameter $u^*$ are the level, the linear, quadratic, etc. components of the C effect. Third, given any $(\beta^*, u^*)$ in which the level or linear component of $u^*$ (or both) are non-zero, construct $\beta^{**}$ and $u^{**}$ so that the residual sum of squares is identical using $(\beta^*, u^*)$ and $(\beta^{**}, u^{**})$ but the penalty $\lambda u^{**\prime} u^{**}$ is smaller than $\lambda u^{*\prime} u^*$. This gives the desired result. We now describe each step in detail.

Step 1: Reparameterize the A and P effects from $\beta$ to $\beta^*$. For the A and P effects, use the reparameterization and symbols given in Holford (1983). If we index the A groups as $i = 1, 2, \ldots, a$, then for observations in the $i^{th}$ A group in the reparameterized design matrix, the column for the linear component has value $i - 0.5a - 0.5$; call this column $A_L$. Similarly, indexing periods as $j = 1, 2, \ldots, p$, for observations in the $j^{th}$ P group in the reparameterized design matrix, the column for the linear component has value $j - 0.5p - 0.5$; call this column $P_L$. For the non-linear components of the A and P effects (which this proof does not use), use the transformations given in Holford (1983). Then the FE $W\beta$ is replaced by $W^*\beta^*$ where $W^* = WK$ for a known invertible matrix $K$ and the new FE parameter vector is $\beta^* = K^{-1}\beta$. Without loss of generality, specify the first three columns of $W^*$ to be $W^{*(0)} = 1_n$ (an n-vector of 1's), $W^{*(1)} = A_L$, and $W^{*(2)} = P_L$, with the remaining columns of $W^*$ being the non-linear components of the A and P effects. The reparameterized FE vector is $\beta^* = (b_0^*, b_{LA}^*, b_{LP}^*, b_C^*)'$, respectively the intercept, linear component of the A effect, linear component of the P effect, and the non-linear components of the two effects gathered into the vector $b_C^*$.

Step 2: Reparameterize the C effect from $u$ to $u^*$. Here, we deviate slightly from Holford (1983) for reasons given below. Reparameterize as $Zu = Z^*u^*$, where $Z^* = Z\Delta$, $u^* = \Delta'u$, and $\Delta$ is an orthogonal matrix (i.e., $\Delta$'s inverse is its transpose so $\Delta'\Delta = \Delta\Delta' = I$, the identity matrix). The rows of $\Delta'$ (which are also the columns of $\Delta$) are the orthogonal polynomials for $c$ levels (Fisher and Yates 1963). The first column of $Z^* = Z\Delta$, is $Z^{*(0)} = k1_n$, for $k = 1/\sqrt{c}$. The second column of $Z^*$ is $Z^{*(1)} = qC_L$, where (as in Holford 1983), the entry in $C_L$ for the k$^{th}$ C group, $k = 1, 2, ..., c$, is $k - 0.5c - 0.5$ and $q = 1/\sqrt{\sum(k - 0.5c - 0.5)^2}$, where the sum inside the square-root symbol is over cohorts $k$. The 3$^{rd}$, 4$^{th}$, etc. columns in $\Delta$ are the quadratic, cubic, etc. orthogonal polynomials, so the 3$^{rd}$, 4$^{th}$, etc. columns of $Z^*$ represent the quadratic, cubic, etc. components of the C effect. The reparameterized C effect is $u^* = (u_0^*, u_L^*, u_C^*)'$, respectively the level (intercept) and linear component of the C effect and the non-linear components of the C effect gathered into a vector $u_C^*$.

We have deviated slightly from Holford (1983) to retain a simple form for the penalty in Equation (6). With the new parameterization, the penalty becomes $\lambda u'u = \lambda u^{*'}\Delta'\Delta u^* = \lambda u^{*'}u^*$ because $\Delta$ was chosen to be an orthogonal matrix. In the new parameterization, the estimates $\beta^*$ and $u$ minimize the penalized residual sum of squares

$$L(\beta^*, u^*) = (y - W^*\beta^* - Z^*u^*)'(y - W^*\beta^* - Z^*u^*) + \lambda u^{*\prime}u^*. \tag{10}$$

Step 3: Construct $\beta^{**}$ and $u^{**}$. Suppose $u_0^* \neq 0$ or $u_L^* \neq 0$ or both. Because $Z^{*(0)} = kW^{*(0)}$, therefore $Z^{*(0)}u_0^* = ku_0^*W^{*(0)}$. Also, from Holford (1983), $C_L = P_L - P_L$, so $Z^{*(1)}u_L^* = q(P_L - A_L)u_L^*$. Using these facts, set $\beta^{**}$ and $u^{**}$ as follows:

$$\beta^{**} = (b_0^* + ku_0^*, b_{LA}^* - qu_L^*, b_{LP}^* + qu_L^*, b_C^*)', \tag{11}$$

$$u^{**} = (0, 0, u_C^*)'. \tag{12}$$

By construction, $(y - W^*\beta^{**} - Z^*u^{**})'(y - W^*\beta^{**} - Z^*u^{**}) = (y - W^*\beta^* - Z^*u^*)'(y - W^*\beta^* - Z^*u^*)$. However, $u^{**\prime}u^{**} = u^{*\prime}u^* - u_0^{*2} - u_L^{*2} < u^{*\prime}u^*$. The result follows.

We have proved that 1-RE-APC models necessarily constrain the estimates of the RE's linear component and intercept to be zero. The following section discusses how 2-RE-APC models are estimated. As we will show, 2-RE-APC models impose implicit constraints to estimate A, P, and C effects as 1-RE-APC models do, but the form of its constraints differs qualitatively from those implied by a 1-RE-APC model.

**5. Constraints in APC Models with Two Random Effects**

We now address the more complicated issue of 2-RE-APC models in which two of three effects are modeled as RE's. Unfortunately, unlike 1-RE models, 2-RE models do not permit an explicit, closed-form restricted likelihood (Hodges 2013), which makes formal proofs difficult. As the best available alternative, we use a particular example, though with little sacrifice of generality: a design with $a = 6$ age groups, $p = 5$ periods, and $c = 10$ cohorts, with P and C modeled as RE's and an outcome measure $y$ on a continuous scale, which we model as normally distributed. After introducing some notation, we use simulated data to show that the estimates of the *linear* components of the P and C effects are determined by the magnitude of the *non-linear* components of the true P and C effects. We then present more technical material explaining how and why this happens.

To simplify the discussion, in our example we assume the true data-generating process has A and P effects that are exactly zero (i.e., no A or P effect), and that only the C effect is non-zero. It is straightforward to produce and explain analogous results when the true A or P effect is non-zero, or when the two RE's are chosen differently.

To introduce some notation, let $\mathbf{W}_a$ be the design matrix for the fixed A effect, which includes the intercept and is parameterized so the A effects sum to zero; let $\beta_a$ be the FE parameter vector. Let $\mathbf{Z}_p$ be the design matrix for the random P effect, with one column for each of the $p$ periods and using the identity parameterization (as in standard software) with parameter $u_p$; and let $\mathbf{Z}_c$ be the design matrix for the random C effect, with one column for each of the $c = a + p - 1 = 6 + 5 - 1 = 10$ cohorts and also using the identity parameterization with parameter $u_c$. Assume the normally distributed errors have variance $\sigma_e^2$ and that the RE variances are $\sigma_p^2$ and $\sigma_c^2$ for P and C respectively. Then given estimates of the three variances, the estimated A, P, and C effects minimize the penalized residual sum of squares (i.e., maximize the penalized log likelihood):

$$(y - \mathbf{W}_a\beta_a - \mathbf{Z}_p u_p - \mathbf{Z}_c u_c)'(y - \mathbf{W}_a\beta_a - \mathbf{Z}_p u_p - \mathbf{Z}_c u_c) + \lambda_p u_p' u_p + \lambda_c u_c' u_c, \quad (13)$$

where $\lambda_p = \sigma_e^2/\sigma_p^2$ and $\lambda_c = \sigma_e^2/\sigma_c^2$.

To simplify the exposition, we reparameterize each of the three effects in Equation (13) using orthogonal polynomials, with each transformation implemented using an orthogonal matrix (i.e., a matrix with the property that its inverse is its transpose), so that the scales of the three design matrices $\mathbf{W}_a$, $\mathbf{Z}_p$, and $\mathbf{Z}_c$ and the three parameter vectors $\beta_a$, $u_p$, and $u_c$ are unchanged and the three variances and the form of the penalty in Equation

(13) are also unchanged. In the following, we will assume that this reparameterization has been made without changing the notation in Equation (13), so that the entries in $\beta_a$ are the intercept and the linear, quadratic, etc. components of the A effect, while the entries in $u_p$ are the level and the linear, quadratic, etc. components of the P effect and analogously for $u_c$ and the C effect.

Now consider a true data-generating mechanism having $\beta_a$ and $u_p$ that are zero vectors (i.e., the true A and P effects are zero), and true C effects of the form $u_c = (0, 1, m, 0, ..., 0)'$ for $m$—the quadratic component of the true C effect—between 0 and 1 inclusive.[4] Recalling that we have reparameterized the cohort effect, the true C effects described by $u_c$ are, in order, the true level (intercept, set to 0), the true linear component, set to 1, the true quadratic component, set to $m$, and then true higher-order components, all set to 0. Setting $m = 0$ means that the true C effect is purely linear; as $m$ grows from 0, C's true linear component remains unchanged while its true non-linear component, to be exact, its quadratic component grows in magnitude. For each value of $m$, we generated 100 datasets by adding $i.i.d.$ errors with true standard deviation 0.01 to this mean, and then we fit an APC model with a fixed A effect and random P and C effects using the lmer function in the lme4 package in R. Appendix C provides the R code we used to generate and analyze these simulated datasets.

Table 5 shows, for each $m$ between 0 and 1, the number of simulated datasets out of 100 for which the lmer function estimated a linear component of the P effect that *was not* shrunk to 0 (or close enough to be effectively 0). Although all datasets used in Table

---

[4] We use this range because using $m < 0$ gives the same results as using $|m|$ and because, as Table 5 shows, for $m > 0.7$, the results do not change.

5 had no true P effect and had C effects with the same true linear component, the estimated *linear* component of the P and C effects changed radically depending on the true *non-linear* component of the C effect, which is determined by $m$. Specifically, when $m = 0$ so the C effect was purely linear, for all 100 simulated datasets lmer gave a zero estimate for $\sigma_c^2$ and a positive estimate for $\sigma_p^2$, so the estimated C effect was shrunk to zero and the true C effect was attributed entirely (and erroneously) to A and P, although in fact the true A and P effects were zero.

[Table 5 about here]

As $m$ grows—that is, as the true non-linear component of C grows in magnitude while the true linear component is held constant—the estimation machinery eventually stops attributing all of C's true linear component to A and P. For example, for $m = 0.30$, six of the 100 datasets gave a positive estimate of $\sigma_c^2$ and a zero estimate of $\sigma_p^2$, so that the P effect's estimated linear component was shrunk to zero (correctly) and C effect's estimated linear component was not. For $m = 0.70$ and larger, for all 100 datasets lmer gave a positive estimate for $\sigma_c^2$ and a zero estimate for $\sigma_p^2$, so the true C effect was correctly attributed to C instead of to A and P. For $m$ between 0.30 and 0.65, the percentage of datasets that erroneously failed to shrink the P effect's linear component to 0 varied but, oddly, not in a steady trend with increasing $m$, as one might expect. Rather, as $m$ increased this percentage went up (as expected), then back down (unexpectedly), before rising finally to 100%.

We have used simulated data for a particular APC design and true data-generating mechanism to show that the estimates of the *linear* components of the P and C effects are determined by the magnitude of the true *non-linear* components of the C effect. We will now explain why this happens, not with the conclusiveness of a mathematical proof but in enough detail to make it clear that the phenomena above could be produced straightforwardly for any choice of two RE's and without assuming that two of the true effects are exactly zero.

To explain the example above, we reiterate how a RE model, in particular our 2-RE-APC model, is estimated. The conventional analysis has two steps. In Step 1, the variances $\sigma_e^2, \sigma_p^2$, and $\sigma_c^2$ are estimated by maximizing the restricted likelihood; in Step 2, the variance ratios $\lambda_p$ and $\lambda_c$ are computed from the estimated variances and the A, P, and C effects are estimated by minimizing the penalized residual sum of squares (13) taking $\lambda_p$ and $\lambda_c$ as given. That is, the specific constraints on the P and C effects imposed by the 2-RE specification are determined by the data in Step 1 and then imposed in Step 2. Although the 2-RE-APC model does not have an explicit, closed-form restricted likelihood, the following discussion uses insights from simpler models that do have closed-form restricted likelihoods (Hodges 2013).

Specifically, in Step 1, the restricted likelihood is computed by regressing out from $y$ the fixed A effect (Searle, Casella, and McCulloch 1992), so that the residuals from the regression of $y$ on A supply all of the information about $\sigma_e^2, \sigma_p^2$, and $\sigma_c^2$ and thus about the ratios $\lambda_p$ and $\lambda_c$. (This is why the restricted likelihood is sometimes called the "residual likelihood".)

This observation allows us to see how the results of Step 1 are strongly affected by the fact that the linear components of the A, P, and C effects are not identified. The column in $\mathbf{Z}_c$ representing the linear component of the C effect, as is well known, is exactly equal to a particular linear combination of the columns in $\mathbf{W}_a$ and $\mathbf{Z}_p$ representing the linear components of the A and P effects respectively. Thus, in computing the restricted likelihood in Step 1, when the regression of $y$ on A is removed from the data $y$, the residual of $y$, which contains the residual of the *true* linear component of C, can be exactly and entirely attributed to the linear component of the P effect *or* to the linear component of the C effect (or to any convex combination of them). That is, the information about the two RE variances $\sigma_p^2$ and $\sigma_c^2$ that is provided to the restricted likelihood by the true C effect's linear component is entirely ambiguous, and that ambiguity is resolved by information in the true non-linear components of the P and C effects.

To understand how the true non-linear component of the C effect produced the phenomena in Table 5, recall that while the linear components of the A, P, and C effects are not identified, their non-linear components *are* identified. In particular, the column in $\mathbf{Z}_c$ representing the quadratic component of the C effect is linearly independent of $\mathbf{W}_a$ and $\mathbf{Z}_p$ though it is not uncorrelated with the columns of those design matrices. Thus in our example with $a = 6, p = 5$, if the column in $\mathbf{Z}_c$ representing the quadratic component of the C effect is regressed on $\mathbf{W}_a$ and $\mathbf{Z}_p$, about 40% of its squared variation is attributed to the fixed A effect and thus lost to the restricted likelihood. Another 6% of the squared variation lies in the column space of $\mathbf{Z}_p$ (which is orthogonal to the column space of $\mathbf{W}_a$); the information in $y$ that lies in this space is again ambiguous as to whether it arose from the P or C effect and thus as to whether it provides information about $\sigma_p^2$ or $\sigma_c^2$. The remaining 54% of the squared variation in $\mathbf{Z}_c$ is unambiguously attributed to the C effect[5]

---

[5] Higher order (cubic, etc.) components of the true C effect, if any, can be partitioned similarly into pieces that Step 1 attributes to the A effect, to the C effect, and ambiguous-

and provides information only about $\sigma_c^2$. Thus, in computing Table 5, as $m$ increased, i.e., as the true C effect's non-linear component increased in magnitude, the restricted likelihood received an increasingly large "signal" that $\sigma_c^2 > 0$, and eventually this resolved the ambiguity of the true cohort effect's linear component, so that it was correctly attributed to the cohort effect and the estimated period effect was correctly shrunk to zero.

This explanation leaves two puzzles. First, when $m = 0$, i.e., when all of the information about $\sigma_p^2$ and $\sigma_c^2$ arises from the true C effect's linear component and that variation could arise entirely from either P or C, why does the P effect "win" for all 100 datasets? And second, why doesn't Table 5 show a nice smooth trend in $m$?

To understand these two puzzles, we examined the profiled restricted likelihood[6], a function of $\sigma_p^2$ and $\sigma_c^2$ that has the same local maxima as the restricted likelihood. For $m = 0$—where the true C effect is exactly linear—the restricted likelihood has two local maxima, one with $\sigma_p^2$ zero and $\sigma_c^2$ positive and the other with $\sigma_p^2$ positive and $\sigma_c^2$ zero. For simulated datasets that we examined, the second local maximum is higher by about 20 natural-log units—a large difference—and thus the global maximum, so the P effect

---

ly to either the P or C effect. Appendix B gives the R code that we used to calculate the fractions given here.

[6] For any pair of values for $\sigma_p^2$ and $\sigma_c^2$, it is easy to derive the value of $\sigma_e^2$ that maximizes the restricted likelihood given that pair $(\sigma_p^2, \sigma_c^2)$. This maximizing value of $\sigma_e^2$ is a function of $(\sigma_p^2, \sigma_c^2)$; substituting it into the restricted likelihood gives the profiled restricted likelihood as a function of $(\sigma_p^2, \sigma_c^2)$.

"wins". This leaves us with another puzzle: *why* does the P effect win? We are unable to answer this question definitively but we conjecture that the P effect wins because it has fewer levels than the C effect ($p$ vs. $p + a - 1$). Because $p < p + a - 1$ in all APC designs, this conjecture cannot be tested for models in which C is an RE. We have examined some examples in which A and P were modeled as RE's and they support the conjecture but examples cannot prove a conjecture.

As $m$ grows from 0, the local maximum with $\sigma_p^2$ zero and $\sigma_c^2$ positive grows higher and the other local maximum, at which $\sigma_p^2$ is positive and $\sigma_c^2$ is zero, becomes lower and shifts its location. As a result, for example, for $m = 0.45$ the latter local maximum occurs for $\sigma_p^2$ and $\sigma_c^2$ with roughly equal values, for which the estimated linear component of the C effect is still shrunk near zero. For these intermediate values of $m$, lmer's code for maximizing the restricted likelihood sometimes finds one local maximum and sometimes finds the other, although the local maxima differ in their values by several log units, so lmer does not necessarily find the global maximum of the restricted likelihood. Thus, the estimates of the variances and the degree of shrinkage imposed on the P and C effects—i.e., the constraint imposed by using the 2-RE-APC model—are determined arbitrarily by the selection of starting values for the numerical maximizer used to maximize the restricted likelihood. As it happens, it is effectively impossible to set starting values for the numerical routine lmer uses to maximize the restricted likelihood, so for intermediate values of $m$ in Table 5, the solution provided by lmer was a haphazard selection from the two local maxima and indeed, because it is impossible to set starting values in lmer's numerical optimizer, lmer could not be used to detect the restricted likelihood's two local maxima even if one knew they existed. SAS's MIXED procedure does allow a user to set starting values for numerically maximizing the restricted likelihood but it uses default starting values of 1 for each variance so an unsuspecting user who relied on these defaults would again be given estimates selected haphazardly by these default starting values.

For $m = 1$—where the true C effect has strong linear and quadratic components—the restricted likelihood has, as far as we can detect, a single maximum with $\sigma_p^2$ zero and $\sigma_c^2$ positive, so the estimated C effect is hardly shrunk at all while the estimated P effect is shrunk to zero, both of which are correct.

In short, all datasets simulated and analyzed for Table 5 had C effects with the same true linear component and yet the estimate of that linear component changed radically depending on the true *non-linear* component of the C effect. What is worse is that the estimate of C's linear component produced by standard software for any given dataset could differ radically depending on the software's choice of starting values for the numerical routine that maximizes the restricted likelihood. Based on the foregoing, we submit that the two-RE analysis is indefensible, and that the problem arises, as always, because the linear components of the three effects are not identified.

## 6. Conclusions and Discussion

In this paper, we proved and illustrated the constraints implicit in random effects age-period-cohort (RE-APC) models. We have shown that such models have greater rank deficiency than fixed-effects (FE) models and documented and illustrated the critical role of the RE assumption in effect estimation using computational and empirical examples. We provided computational results and a mathematical proof to support our conclusion that the RE mechanism serves to supply the critical yet implicit constraints to identify these otherwise unidentified models. It is not surprising, then, that the largely arbitrary implicit constraints do not just shrink the magnitude of the effects modeled as random, as many researchers believe; rather, these constraints *dictate* estimation results and substantive conclusions.

It is interesting, but not noted in previous literature, that the specific constraints implicit in the RE specification are not the same for RE-APC models with one RE (1-RE-APC model) and with two RE's (2-RE-APC model). Specifically, for 1-RE-APC models, O'Brien (2017) noted that the estimated linear trend in the RE effect appears to be zero in data analyses. We have carried forward this observation with computational results and a mathematical proof that using the 1-RE-APC has the effect of forcing to zero the estimated linear component and intercept (or level) of the effect modeled as random. For example, researchers should be aware that when the C effects are modeled as random in an APC analysis, this choice pre-determines that both the linear component and the level of the random C intercepts will be estimated as zero. Such an assumption usually lacks theoretical justification; this is especially concerning when the assumption is implicit in the most obscure part of the model specification.

For 2-RE-APC models, Bell and Jones (2018) made the important observation that even for the same data, changing the number of A, P, and thus C groups changes the estimation results markedly, which should not occur with identified models. We advanced this literature by showing that the constraints implicit in a 2-RE-APC model also depend on the size of the true non-linear components of the RE's. That is, even when two researchers analyze their data using the same number of levels for each effect, the esti-mated linear trend in an effect modeled as random may differ if the *non-linear* components of that effect differ in magnitude.

The identification problem in RE-APC models ought not be confused with a violation of assumptions in a statistical model. Indeed, one assumption in a standard linear regression model is that the observations are independently distributed. RE models are often used when the independence assumption is violated. For APC models, it has been argued that this impendence assumption is violated in that individuals are nested within time periods and birth cohorts (see, e.g., Yang and Land 2006). If this were simply a violation of the independence assumption and RE-APC models were identifiable without constraints, then estimation results of RE-APC models with different RE choices should be similar to each other and to those from an FE-APC model. To the contrary, we and others have shown that estimates of the A, P, and C effects depend strongly on which effect is considered random and on the size of the non-linear component of each effect that is modeled as random. That is, although it is not unreasonable to assume a hierarchical structure in APC data, researchers should be aware that the RE specification largely predetermines certain components of the estimated A, P, and C effects.

The debate surrounding APC models is decades old. Many methods have been proposed, some relying on technical assumptions (see, e.g., Robertson and Boyle 1998; Yang et al. 2008) and others based on social theory (Fosse and Winship forthcoming; Mason and Fienberg 1985; O'Brien 2000) or hypothesized mechanisms or proxy variables through which causal factors affect the outcome measures (Heckman and Robb 1985; Winship and Harding 2008). In the end, what researchers can hope to accomplish depends on what they intend the statistical method to do (Luo 2013b). In many demograph-ic and social sciences applications, statistical models are useful for summarizing information in the data in a concise way. Our investigation of the dramatically different estimates from RE-APC models strongly suggests that researchers should not only stop using RE models as an attempt to recover the "true" A, P, and C effects without a strong theoretical account but also should prioritize critically examining the conceptual framework that motivates the undertaking to estimate effects that have such particular forms.

**References**

Bell, Andrew and Kelvyn Jones. 2018. "The Hierarchical Age–Period–Cohort Model: Why Does It Find the Results That It Finds?" *Quality & Quantity* 52(2):783–99.

Davis, James A. 2004. "Did Growing Up in the 1960s Leave a Permanent Mark on Attitudes and Values?Evidence from the General Social Survey." *Public Opinion Quarterly* 68(2):161–83.

Davis, James A. 2013. "A Generation of Attitude Trends among US Householders as Measured in the NORC General Social Survey 1972–2010." *Social Science Research* 42(3):571–83.

Fienberg, Stephen E. 2013. "Cohort Analysis' Unholy Quest: A Discussion." *Demography* 50(6):1981–84.

Fienberg, Stephen E. and William M. Mason. 1979. "Identification and Estimation of Age-Period-Cohort Models in the Analysis of Discrete Archival Data." *Sociological Methodology* 10:1–67.

Fisher, Ronald Aylmer and Frank Yates. 1963. *Statistical Tables for Biological, Agricultural and Medical Research, Edited by R.A. Fisher and F. Yates. 6th Ed.* Edinburgh: Oliver and Boyd.

Fosse, Ethan and Christopher Winship. Forthcoming. "Bounding Analyses of Age-Period-Cohort Effects." *Demography*.

Gauchat, Gordon. 2012. "Politicization of Science in the Public Sphere: A Study of Public Trust in the United States, 1974 to 2010." *American Sociological Review* 77(2):167–87.

Gelman, Andrew and Jennifer Hill. 2007. *Data Analysis Using Regression and Multilevel/Hierarchical Models*. Cambridge University Press.

Glass, Jennifer. 1992. "Housewives and Employed Wives: Demographic and Attitudinal Change, 1972-1986." *Journal of Marriage and Family* 54(3):559–69.

Hagerty, Michael R. and Ruut Veenhoven. 2003. "Wealth and Happiness Revisited – Growing National Income Does Go with Greater Happiness." *Social Indicators Research* 64(1):1–27.

Heckman, James and Richard Robb. 1985. "Using Longitudinal Data to Estimate Age, Period and Cohort Effects in Earnings Equations." Pp. 137–50 in *Cohort Analysis in Social Research*, edited by W. M. Mason and S. E. Fienberg. Springer New York.

Hodges, James S. 2013. *Richly Parameterized Linear Models: Additive, Time Series, and Spatial Models Using Random Effects*. CRC Press.

Holford, Theodore R. 1983. "The Estimation of Age, Period and Cohort Effects for Vital Rates." *Biometrics* 39(2):311–24.

Hout, Michael and Claude S. Fischer. 2014. "Explaining Why More Americans Have No Religious Preference: Political Backlash and Generational Succession, 1987-2012." *Sociological Science* 1:423–47.

Jaacks, Lindsay M., Penny Gordon-Larsen, Elizabeth J. Mayer-Davis, Linda S. Adair, and Barry Popkin. 2013. "Age, Period and Cohort Effects on Adult Body Mass Index and Overweight from 1991 to 2009 in China: The China Health and Nutrition Survey." *International Journal of Epidemiology* 42(3):828–37.

Kupper, Lawrence L., Joseph M. Janis, Azza Karmous, and Bernard G. Greenberg. 1985. "Statistical Age-Period-Cohort Analysis: A Review and Critique." *Journal of Chronic Diseases* 38(10):811–30.

Luo, Liying. 2013a. "Assessing Validity and Application Scope of the Intrinsic Estimator Approach to the Age-Period-Cohort Problem." *Demography* 50(6):1945–67.

Luo, Liying. 2013b. "Paradigm Shift in Age-Period-Cohort Analysis: A Response to Yang and Land, O'Brien, Held and Riebler, and Fienberg." *Demography* 50(6):1985–88.

Luo, Liying and James S. Hodges. 2016. "Block Constraints in Age–Period–Cohort Models with Unequal-Width Intervals." *Sociological Methods & Research* 45(4):700–726.

Luo, Liying, James Hodges, Christopher Winship, and Daniel Powers. 2016. "The Sensitivity of the Intrinsic Estimator to Coding Schemes: Comment on Yang, Schulhofer-Wohl, Fu, and Land." *American Journal of Sociology* 122(3):930–61.

Mason, William M. and Stephen E. Fienberg. 1985. "Introduction: Beyond the Identification Problem." Pp. 1–8 in *Cohort Analysis in Social Research*. Springer, New York, NY.

Masters, Ryan K., Robert A. Hummer, and Daniel A. Powers. 2012. "Educational Differences in U.S. Adult Mortality: A Cohort Perspective." *American Sociological Review* 77(4):548–72.

O'Brien, Robert M. 2000. "Age Period Cohort Characteristic Models." *Social Science Research* 29(1):123–39.

O'Brien, Robert M. 2013. "Comment of Liying Luo's Article, 'Assessing Validity and Application Scope of the Intrinsic Estimator Approach to the Age-Period-Cohort Problem.'" *Demography* 50(6):1973–75; discussion 1985-1988.

O'Brien, Robert M. 2017. "Mixed Models, Linear Dependency, and Identification in Age-Period-Cohort Models." *Statistics in Medicine* 36(16):2590–2600.

Pampel, Fred C. and Lori M. Hunter. 2012. "Cohort Change, Diffusion, and Support for Environmental Spending in the United States." *American Journal of Sociology* 118(2):420–48.

Raudenbush, Stephen W. and Anthony S. Bryk. 2002. *Hierarchical Linear Models: Applications and Data Analysis Methods*. SAGE.

Robertson, Chris and Peter Boyle. 1998. "Age–Period–Cohort Analysis of Chronic Disease Rates. I: Modelling Approach." *Statistics in Medicine* 17(12):1305–23.

Robinson, Robert V. and Elton F. Jackson. 2001. "Is Trust in Others Declining in America? An Age–Period–Cohort Analysis." *Social Science Research* 30(1):117–45.

Rodgers, Willard L. 1982. "Estimable Functions of Age, Period, and Cohort Effects." *American Sociological Review* 47(6):774–87.

Schwadel, Philip. 2010. "Age, Period, and Cohort Effects on U.S. Religious Service Attendance: The Declining Impact of Sex, Southern Residence, and Catholic Affiliation." *Sociology of Religion* 71(1):2–24.

Searle, Shayle R., George Casella, and Charles E. McCulloch. 1992. *Variance Components*. Wiley.

Shu, Xiaoling and Yifei Zhu. 2012. "Uneven Transitions: Period- and Cohort-Related Changes in Gender Attitudes in China, 1995–2007." *Social Science Research* 41(5):1100–1115.

Smets, Kaat and Anja Neundorf. 2014. "The Hierarchies of Age-Period-Cohort Research: Political Context and the Development of Generational Turnout Patterns." *Electoral Studies* 33:41–51.

Smith, Tom W. 1990. "Liberal and Conservative Trends in the United States Since World War II." *The Public Opinion Quarterly* 54(4):479–507.

Winship, Christopher and David J. Harding. 2008. "A Mechanism-Based Approach to the Identification of Age–Period–Cohort Models." *Sociological Methods & Research* 36(3):362–401.

Yang, Yang. 2008. "Social Inequalities in Happiness in the United States, 1972 to 2004: An Age-Period-Cohort Analysis." *American Sociological Review* 73(2):204–26.

Yang, Yang, Sam Schulhofer-Wohl, Wenjiang J. Fu, and Kenneth C. Land. 2008. "The Intrinsic Estimator for Age-Period-Cohort Analysis: What It Is and How to Use It." *American Journal of Sociology* 113(6):1697–1736.

Table 1. Design Matrix of an FE-APC Model.

| Intercept | Age | | Period | | Cohort | | | |
|---|---|---|---|---|---|---|---|---|
| | $\alpha_1$ | $\alpha_2$ | $\beta_1$ | $\beta_2$ | $\gamma_1$ | $\gamma_2$ | $\gamma_3$ | $\gamma_4$ |
| 1 | 1 | 0 | 1 | 0 | 0 | 0 | 1 | 0 |
| 1 | 1 | 0 | 0 | 1 | 0 | 0 | 0 | 1 |
| 1 | 1 | 0 | -1 | -1 | -1 | -1 | -1 | -1 |
| 1 | 0 | 1 | 1 | 0 | 0 | 1 | 0 | 0 |
| 1 | 0 | 1 | 0 | 1 | 0 | 0 | 1 | 0 |
| 1 | 0 | 1 | -1 | -1 | 0 | 0 | 0 | 1 |
| 1 | -1 | -1 | 1 | 0 | 1 | 0 | 0 | 0 |
| 1 | -1 | -1 | 0 | 1 | 0 | 1 | 0 | 0 |
| 1 | -1 | -1 | -1 | -1 | 0 | 0 | 1 | 0 |

Table 2. Design Matrix of a 1-RE-APC Model.

| Fixed Effects | | | | | Random Effects | | | | |
|---|---|---|---|---|---|---|---|---|---|
| Intercept | Age | | Period | | Cohort | | | | |
|  | $\alpha_1$ | $\alpha_2$ | $\beta_1$ | $\beta_2$ | $\gamma_1$ | $\gamma_2$ | $\gamma_3$ | $\gamma_4$ | $\gamma_5$ |
| 1 | 1 | 0 | 1 | 0 | 0 | 0 | 1 | 0 | 0 |
| 1 | 1 | 0 | 0 | 1 | 0 | 0 | 0 | 1 | 0 |
| 1 | 1 | 0 | -1 | -1 | 0 | 0 | 0 | 0 | 1 |
| 1 | 0 | 1 | 1 | 0 | 0 | 1 | 0 | 0 | 0 |
| 1 | 0 | 1 | 0 | 1 | 0 | 0 | 1 | 0 | 0 |
| 1 | 0 | 1 | -1 | -1 | 0 | 0 | 0 | 1 | 0 |
| 1 | -1 | -1 | 1 | 0 | 1 | 0 | 0 | 0 | 0 |
| 1 | -1 | -1 | 0 | 1 | 0 | 1 | 0 | 0 | 0 |
| 1 | -1 | -1 | -1 | -1 | 0 | 0 | 1 | 0 | 0 |

Table 3. Design Matrix of a 2-RE-APC Model.

| Fixed Effects | | | Random Effects | | | | | | | |
|---|---|---|---|---|---|---|---|---|---|---|
| Intercept | Age | | Period | | | Cohort | | | | |
| | $\alpha_1$ | $\alpha_2$ | $\beta_1$ | $\beta_2$ | $\beta_3$ | $\gamma_1$ | $\gamma_2$ | $\gamma_3$ | $\gamma_4$ | $\gamma_5$ |
| 1 | 1 | 0 | 1 | 0 | 0 | 0 | 0 | 1 | 0 | 0 |
| 1 | 1 | 0 | 0 | 1 | 0 | 0 | 0 | 0 | 1 | 0 |
| 1 | 1 | 0 | 0 | 0 | 1 | 0 | 0 | 0 | 0 | 1 |
| 1 | 0 | 1 | 1 | 0 | 0 | 0 | 1 | 0 | 0 | 0 |
| 1 | 0 | 1 | 0 | 1 | 0 | 0 | 0 | 1 | 0 | 0 |
| 1 | 0 | 1 | 0 | 0 | 1 | 0 | 0 | 0 | 1 | 0 |
| 1 | -1 | -1 | 1 | 0 | 0 | 1 | 0 | 0 | 0 | 0 |
| 1 | -1 | -1 | 0 | 1 | 0 | 0 | 1 | 0 | 0 | 0 |
| 1 | -1 | -1 | 0 | 0 | 1 | 0 | 0 | 1 | 0 | 0 |

Table 4. Computational Results: Estimates of the Linear Component and Random Intercept in Largest Absolute Value.

| Design | a | p | λ | Linear Component | Random Intercept |
|---|---|---|---|---|---|
| 1 | 3 | 3 | 0.001 | 0.000 | 0.000 |
| 2 | 3 | 3 | 0.010 | 0.000 | 0.000 |
| 3 | 3 | 3 | 0.100 | 0.000 | 0.000 |
| 4 | 3 | 3 | 1 | 0.000 | 0.000 |
| 5 | 3 | 3 | 10 | 0.000 | 0.000 |
| 6 | 3 | 3 | 100 | 0.000 | 0.000 |
| 7 | 3 | 3 | 1000 | 0.000 | 0.000 |
| 8 | 3 | 4 | 0.001 | 0.000 | 0.000 |
| 9 | 3 | 4 | 0.010 | 0.000 | 0.000 |
| … | | | … | | |
| 195 | 3 | 30 | 100 | | |
| 196 | 3 | 30 | 1000 | | |
| 197 | 4 | 3 | 0.001 | 0.000 | 0.000 |
| 198 | 4 | 3 | 0.010 | 0.000 | 0.000 |
| 199 | 4 | 3 | 0.100 | 0.000 | 0.000 |
| 200 | 4 | 3 | 1 | 0.000 | 0.000 |
| 201 | 4 | 3 | 10 | 0.000 | 0.000 |
| 202 | 4 | 3 | 100 | 0.000 | 0.000 |
| 203 | 4 | 3 | 1000 | 0.000 | 0.000 |
| 204 | 4 | 4 | 0.001 | 0.000 | 0.000 |
| 205 | 4 | 4 | 0.010 | 0.000 | 0.000 |
| … | | | … | | |
| 5488 | 30 | 30 | 1000 | 0.000 | 0.000 |

Note: a denotes the number of age groups in an APC design; p denotes the number of periods; λ denotes the error and RE variance ratio as defined in Equation (8).

Table 5. Number of Datasets out of 100 for which the Estimated Period Linear Component Is Shrunk Effectively to Zero.

| $m$ | # of datasets with zero period linear component |
|---|---|
| 0 | 0 |
| 0.05 | 0 |
| 0.1 | 0 |
| 0.15 | 0 |
| 0.2 | 0 |
| 0.25 | 0 |
| 0.3 | 6 |
| 0.35 | 40 |
| 0.4 | 80 |
| 0.45 | 46 |
| 0.5 | 37 |
| 0.55 | 6 |
| 0.6 | 2 |
| 0.65 | 90 |
| 0.7 | 100 |
| 0.75 | 100 |
| 0.8 | 100 |
| 0.85 | 100 |
| 0.9 | 100 |
| 0.95 | 100 |
| 1 | 100 |

Note: $m$ denotes the quadratic component of the true C effects in the data-generating mechanims.

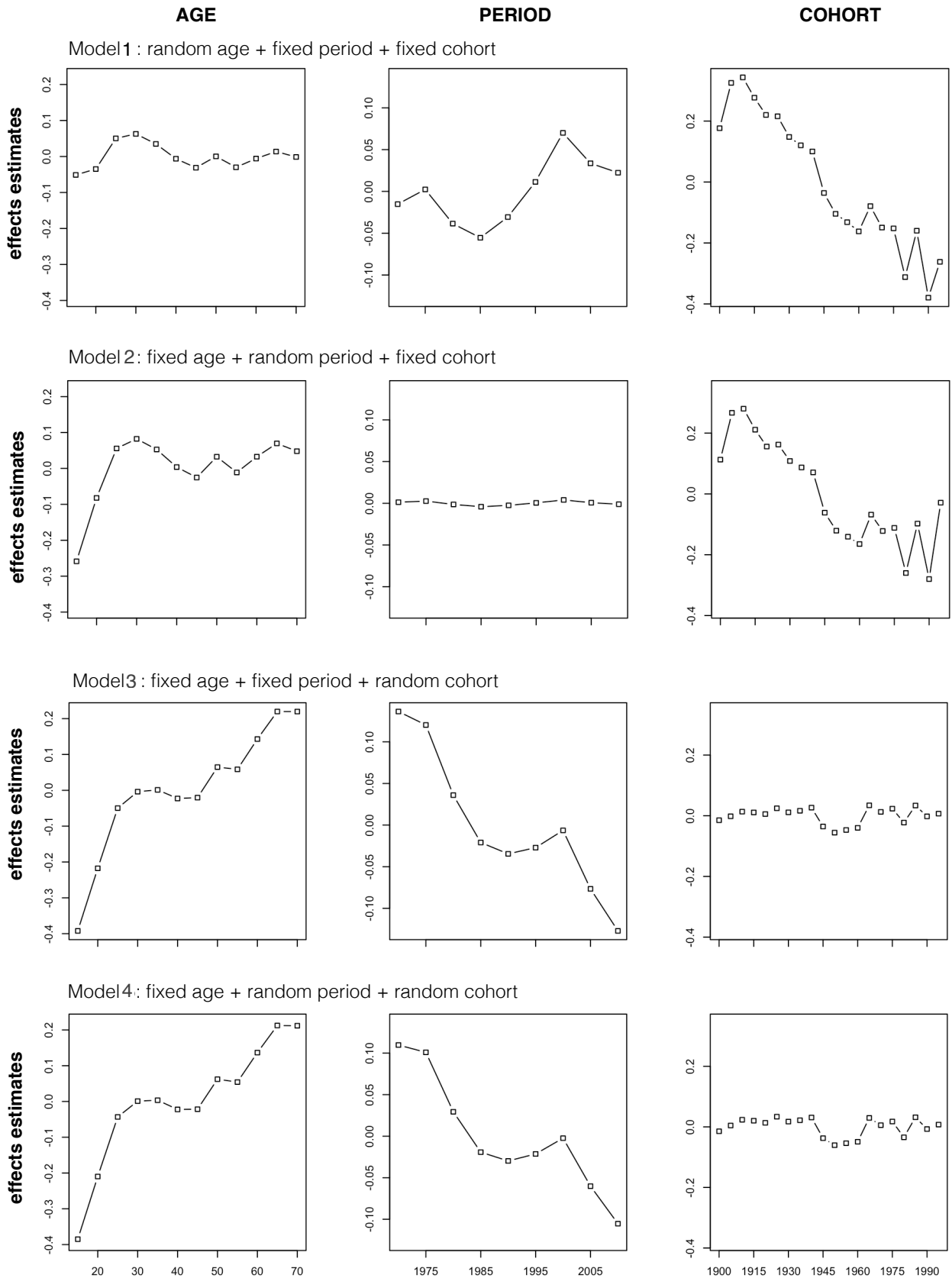

**Figure 1. Sensitivity of the Estimates to RE Choice in Six RE-APC Models for the GSS Happiness Data.**

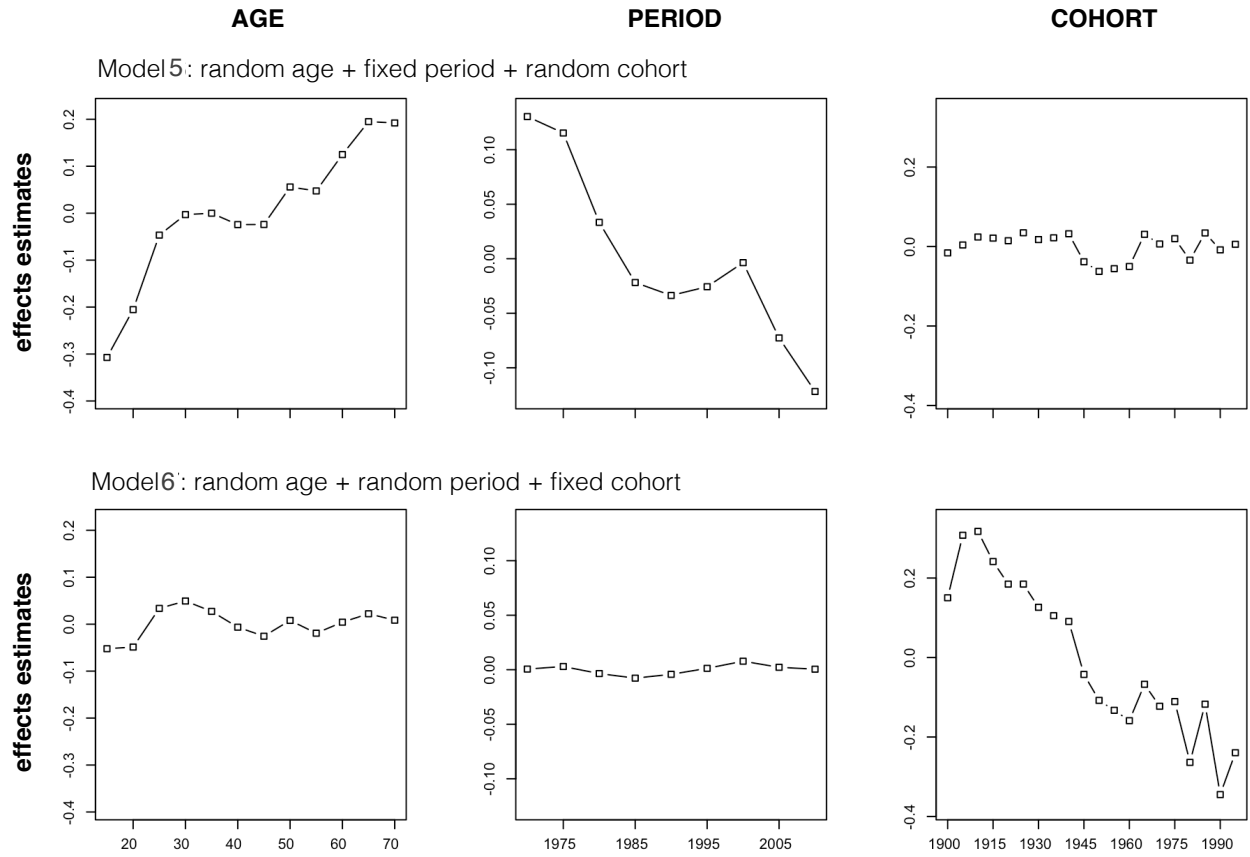

**Figure 1 conti.**

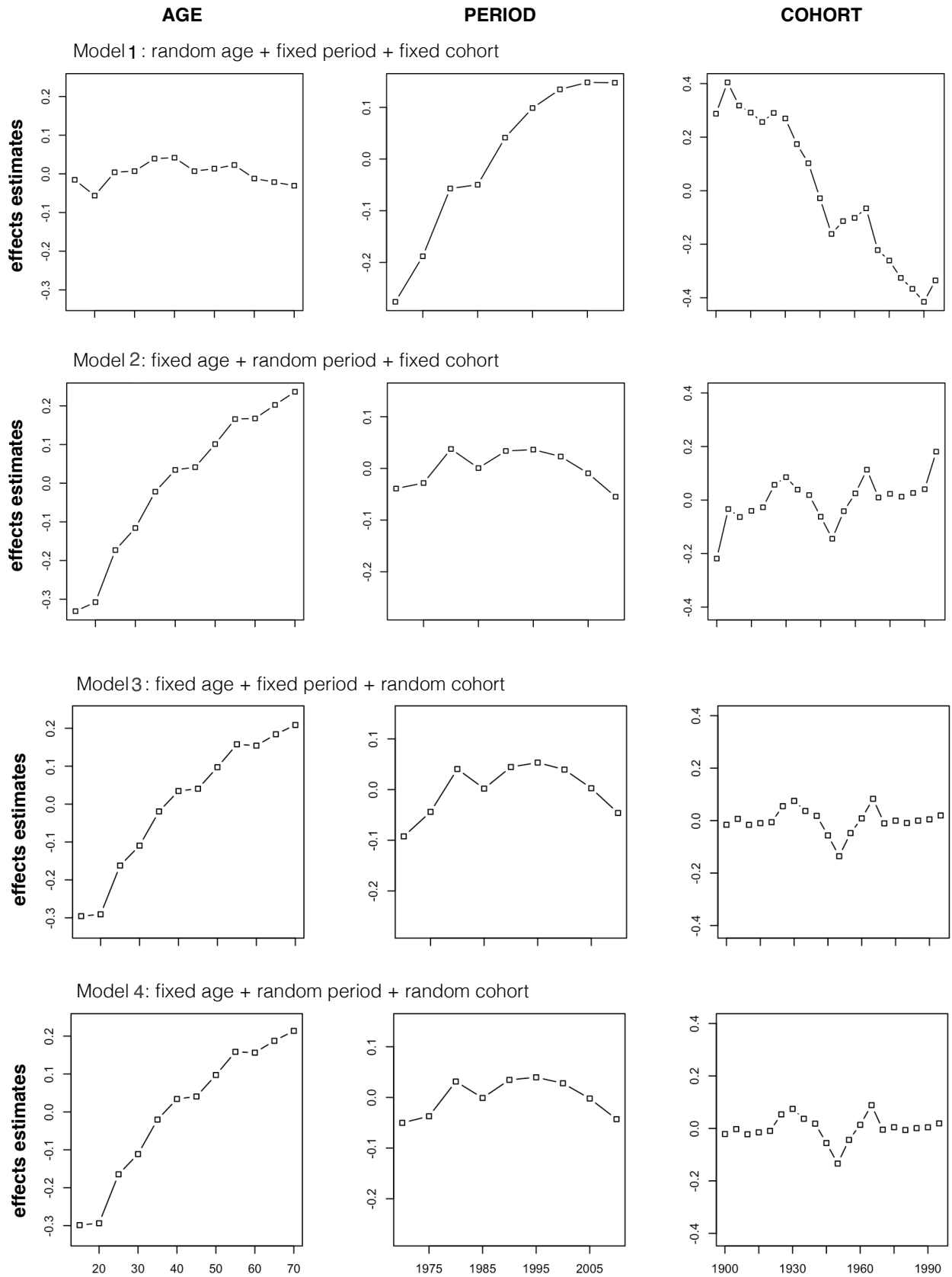

**Figure 2. Sensitivity of the Estimates to RE Choice in Six RE-APC Models for the GSS Conservative-Liberal Political Views Data.**

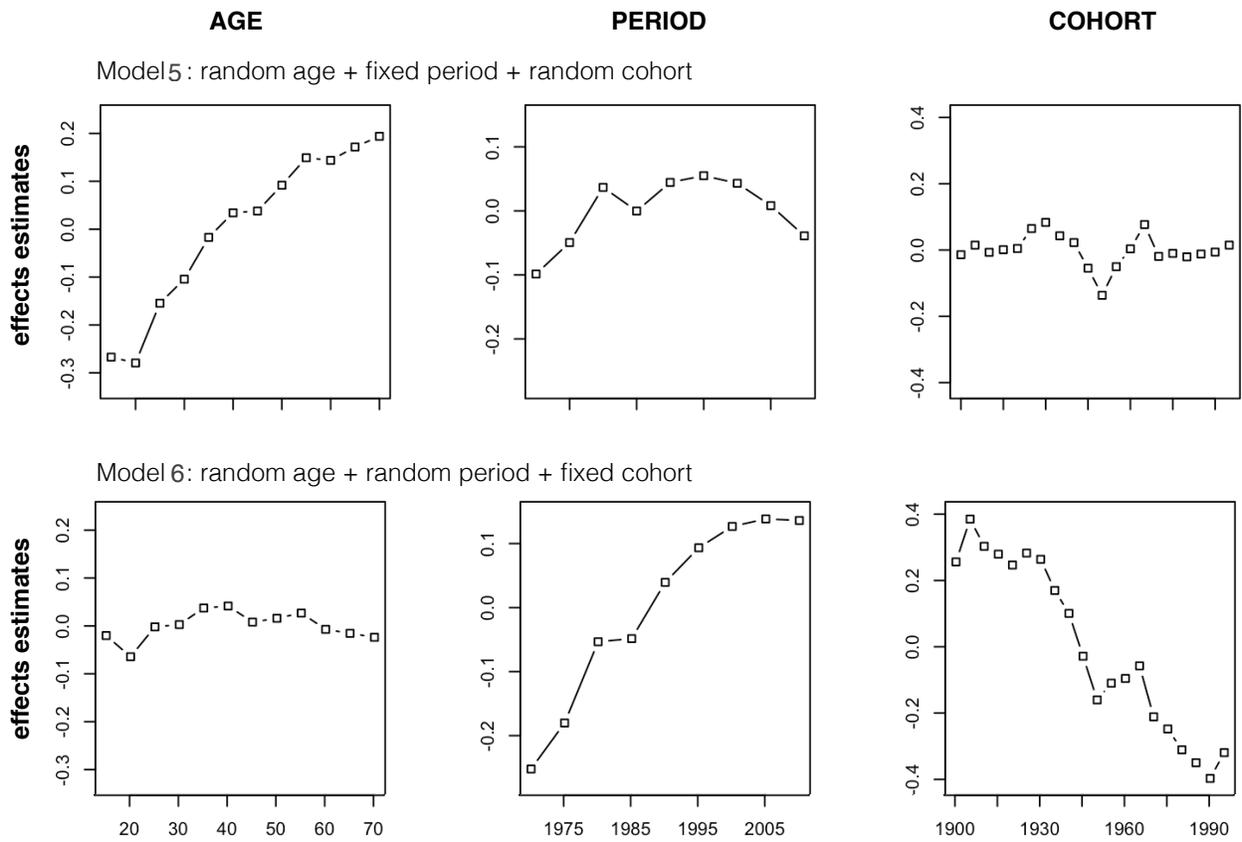

**Figure 2 conti.**

Table 1A. Age, Period, and Cohort Effects Estimates on Happiness from Six RE-APC Models, the GSS 1974-2014.

| | Group | Model 1 | Model 2 | Model 3 | Model 4 | Model 5 | Model 6 |
|---|---|---|---|---|---|---|---|
| **Age Effects** | 1 | -0.051 | -0.259 | -0.392 | -0.385 | -0.307 | -0.052 |
| | 2 | -0.035 | -0.082 | -0.218 | -0.210 | -0.205 | -0.049 |
| | 3 | 0.050 | 0.056 | -0.050 | -0.043 | -0.047 | 0.034 |
| | 4 | 0.063 | 0.082 | -0.004 | 0.001 | -0.003 | 0.049 |
| | 5 | 0.035 | 0.053 | 0.001 | 0.004 | 0.000 | 0.027 |
| | 6 | -0.006 | 0.004 | -0.023 | -0.022 | -0.024 | -0.006 |
| | 7 | -0.031 | -0.025 | -0.020 | -0.022 | -0.024 | -0.026 |
| | 8 | 0.000 | 0.033 | 0.065 | 0.062 | 0.056 | 0.008 |
| | 9 | -0.030 | -0.011 | 0.058 | 0.054 | 0.047 | -0.019 |
| | 10 | -0.006 | 0.033 | 0.143 | 0.137 | 0.125 | 0.004 |
| | 11 | 0.014 | 0.070 | 0.220 | 0.212 | 0.195 | 0.022 |
| | 12 | -0.001 | 0.048 | 0.220 | 0.212 | 0.192 | 0.009 |
| **Period Effects** | 1 | -0.015 | 0.001 | 0.136 | 0.110 | 0.130 | 0.001 |
| | 2 | 0.002 | 0.003 | 0.120 | 0.101 | 0.115 | 0.003 |
| | 3 | -0.039 | -0.001 | 0.036 | 0.029 | 0.033 | -0.004 |
| | 4 | -0.055 | -0.004 | -0.021 | -0.019 | -0.022 | -0.008 |
| | 5 | -0.031 | -0.002 | -0.035 | -0.030 | -0.034 | -0.004 |
| | 6 | 0.011 | 0.000 | -0.027 | -0.021 | -0.026 | 0.001 |
| | 7 | 0.070 | 0.004 | -0.006 | -0.002 | -0.004 | 0.008 |
| | 8 | 0.034 | 0.001 | -0.077 | -0.060 | -0.073 | 0.002 |
| | 9 | 0.022 | -0.001 | -0.127 | -0.105 | -0.122 | 0.001 |
| **Cohort Effects** | 1 | 0.177 | 0.113 | -0.015 | -0.015 | -0.016 | 0.150 |
| | 2 | 0.325 | 0.267 | -0.002 | 0.005 | 0.004 | 0.308 |
| | 3 | 0.344 | 0.280 | 0.014 | 0.024 | 0.024 | 0.317 |
| | 4 | 0.277 | 0.211 | 0.011 | 0.020 | 0.021 | 0.242 |
| | 5 | 0.220 | 0.156 | 0.006 | 0.013 | 0.015 | 0.185 |
| | 6 | 0.216 | 0.162 | 0.025 | 0.034 | 0.035 | 0.185 |
| | 7 | 0.148 | 0.108 | 0.011 | 0.017 | 0.017 | 0.126 |
| | 8 | 0.120 | 0.087 | 0.016 | 0.022 | 0.022 | 0.105 |
| | 9 | 0.100 | 0.071 | 0.027 | 0.031 | 0.032 | 0.091 |
| | 10 | -0.036 | -0.062 | -0.035 | -0.037 | -0.038 | -0.043 |
| | 11 | -0.104 | -0.121 | -0.056 | -0.060 | -0.063 | -0.108 |
| | 12 | -0.132 | -0.141 | -0.047 | -0.054 | -0.056 | -0.133 |
| | 13 | -0.162 | -0.165 | -0.040 | -0.049 | -0.050 | -0.159 |
| | 14 | -0.079 | -0.068 | 0.034 | 0.030 | 0.031 | -0.067 |
| | 15 | -0.149 | -0.122 | 0.013 | 0.006 | 0.006 | -0.123 |
| | 16 | -0.152 | -0.112 | 0.023 | 0.018 | 0.020 | -0.111 |
| | 17 | -0.312 | -0.260 | -0.023 | -0.034 | -0.034 | -0.264 |
| | 18 | -0.160 | -0.098 | 0.034 | 0.032 | 0.034 | -0.117 |
| | 19 | -0.379 | -0.280 | -0.002 | -0.007 | -0.008 | -0.345 |
| | 20 | -0.262 | -0.029 | 0.007 | 0.008 | 0.006 | -0.240 |

Note: Model 1: random A effect and fixed P and C effects; Model 2: random P effect and fixed A and C effects; Model 3: random C effect and fixed A and P effects; Model 4: fixed A effect and random P and C effects; Model 5: fixed P effect and random A and C effects; Model 6: fixed C effects and random A and P effects; in all models, fixed effects were estimated using the sum-to-zero constraint.

Table 2A. Age, Period, and Cohort Effects Estimates on Conservative-Liberal Political Views from Six RE-APC Models, the GSS 1974-2014.

|  | Group | Model 1 | Model 2 | Model 3 | Model 4 | Model 5 | Model 6 |
|---|---|---|---|---|---|---|---|
| **Age Effects** | 1 | -0.015 | -0.331 | -0.296 | -0.299 | -0.267 | -0.020 |
| | 2 | -0.056 | -0.308 | -0.291 | -0.294 | -0.279 | -0.064 |
| | 3 | 0.004 | -0.173 | -0.162 | -0.165 | -0.155 | -0.002 |
| | 4 | 0.007 | -0.116 | -0.110 | -0.111 | -0.104 | 0.003 |
| | 5 | 0.039 | -0.022 | -0.019 | -0.020 | -0.017 | 0.037 |
| | 6 | 0.042 | 0.034 | 0.035 | 0.034 | 0.034 | 0.042 |
| | 7 | 0.007 | 0.041 | 0.041 | 0.041 | 0.038 | 0.008 |
| | 8 | 0.013 | 0.101 | 0.097 | 0.097 | 0.092 | 0.016 |
| | 9 | 0.023 | 0.166 | 0.158 | 0.158 | 0.149 | 0.027 |
| | 10 | -0.012 | 0.167 | 0.154 | 0.156 | 0.144 | -0.007 |
| | 11 | -0.021 | 0.202 | 0.184 | 0.188 | 0.172 | -0.015 |
| | 12 | -0.031 | 0.236 | 0.209 | 0.214 | 0.194 | -0.024 |
| **Period Effects** | 1 | -0.276 | -0.039 | -0.092 | -0.050 | -0.099 | -0.252 |
| | 2 | -0.188 | -0.028 | -0.044 | -0.037 | -0.049 | -0.180 |
| | 3 | -0.057 | 0.038 | 0.041 | 0.031 | 0.037 | -0.053 |
| | 4 | -0.050 | 0.001 | 0.002 | -0.001 | 0.000 | -0.048 |
| | 5 | 0.041 | 0.034 | 0.045 | 0.035 | 0.044 | 0.039 |
| | 6 | 0.099 | 0.036 | 0.053 | 0.040 | 0.055 | 0.093 |
| | 7 | 0.135 | 0.023 | 0.039 | 0.028 | 0.043 | 0.127 |
| | 8 | 0.148 | -0.009 | 0.003 | -0.002 | 0.008 | 0.139 |
| | 9 | 0.148 | -0.055 | -0.046 | -0.043 | -0.039 | 0.136 |
| **Cohort Effects** | 1 | 0.288 | -0.219 | -0.016 | -0.021 | -0.014 | 0.256 |
| | 2 | 0.405 | -0.034 | 0.007 | -0.003 | 0.015 | 0.385 |
| | 3 | 0.318 | -0.063 | -0.016 | -0.022 | -0.007 | 0.303 |
| | 4 | 0.292 | -0.040 | -0.010 | -0.015 | 0.001 | 0.279 |
| | 5 | 0.257 | -0.027 | -0.006 | -0.010 | 0.005 | 0.247 |
| | 6 | 0.291 | 0.057 | 0.055 | 0.054 | 0.065 | 0.283 |
| | 7 | 0.270 | 0.086 | 0.075 | 0.075 | 0.083 | 0.264 |
| | 8 | 0.174 | 0.039 | 0.037 | 0.037 | 0.043 | 0.170 |
| | 9 | 0.103 | 0.019 | 0.018 | 0.018 | 0.023 | 0.101 |
| | 10 | -0.028 | -0.062 | -0.057 | -0.056 | -0.055 | -0.029 |
| | 11 | -0.162 | -0.144 | -0.136 | -0.134 | -0.136 | -0.160 |
| | 12 | -0.113 | -0.042 | -0.048 | -0.044 | -0.050 | -0.110 |
| | 13 | -0.102 | 0.025 | 0.009 | 0.014 | 0.004 | -0.096 |
| | 14 | -0.066 | 0.113 | 0.083 | 0.089 | 0.077 | -0.058 |
| | 15 | -0.222 | 0.009 | -0.010 | -0.005 | -0.019 | -0.211 |
| | 16 | -0.261 | 0.023 | 0.000 | 0.005 | -0.010 | -0.248 |
| | 17 | -0.326 | 0.013 | -0.009 | -0.006 | -0.020 | -0.311 |
| | 18 | -0.367 | 0.026 | 0.000 | 0.001 | -0.012 | -0.350 |
| | 19 | -0.415 | 0.040 | 0.005 | 0.004 | -0.006 | -0.397 |
| | 20 | -0.335 | 0.181 | 0.020 | 0.019 | 0.015 | -0.319 |

Note: Model 1: random A effect and fixed P and C effects; Model 2: random P effect and fixed A and C effects; Model 3: random C effect and fixed A and P effects; Model 4: fixed A effect and random P and C effects; Model 5: fixed P effect and random A and C effects; Model 6: fixed C effects and random A and P effects; in all models, fixed effects were estimated using the sum-to-zero constraint.

**Appendix A: R code for the computational results for 1-RE-APC models**

For all APC designs with $a$ (number of age groups) and $p$ (number of periods) in given ranges (selected by the user) and for all variance ratios $\lambda = \sigma_e^2 / \sigma_c^2$ with $\lambda$ in 0.001 to 1000 (or $log_{10}\lambda$ in -3 to 3), show that the intercept and linear component of the fitted cohort effect is necessarily 0 for all datasets $y$. A dataset $y$ is a column vector of length $a \times p$, sorted first by age group and then by period within age group (i.e., the period index varies fastest)

```
# Specify ranges to consider for a, p, and log_10 λ
# number of age groups
a <- 3:30
na <- length(a)
# number of periods
p <- 3:30
np <- length(p)
# number of lambda values, i.e., variance ratios
lam <- 10^(-3:3)
nl <- length(lam)

# Set up the matrix apmax to hold results.  In apmax,
# columns 1, 2, and 3 are the number of age groups, number of periods, and λ, respectively.
# Columns 4 and 5 are, for the cohort intercept and linear component estimates, respectively,
# the maximum of the absolute values of the coefficients of the data y.
apmax <- matrix(NA,na*np*nl,5)

# Loop 1, over number of age groups
for(i in 1:na){

# Loop 2, over number of periods
for(j in 1:np){

# design matrix columns for age, then period
   Xa <- contr.sum(n=a[i]) %x% as.matrix(rep(1,p[j]))
   Xp <- as.matrix(rep(1,a[i])) %x% contr.sum(n=p[j])
# column bind with a column of 1s to give the fixed effect design matrix W
   W <- cbind(matrix(1,a[i]*p[j],1),Xa,Xp)
# design matrix for cohort random effect
   Z <- matrix(0,a[i]*p[j],a[i]+p[j]-1)
   for(it in 1:a[i]){for(jt in 1:p[j]){
      Z[(it-1)*p[j]+jt,a[i] + jt - it] <- 1  }}
# column bind Z with W to give Q; compute Q'Q
   Q <- cbind(W,Z)
   QpQ <- t(Q) %*% Q
# Set up the penalty matrix D for cohort
   D <- diag(c(rep(0,a[i]+p[j]-1),rep(1,a[i]+p[j]-1)))
# set up alf and bet, which give the intercept and linear cohort estimates respectively
   alf <- as.matrix(c(rep(0,a[i]+p[j]-1),rep(1/(a[i]+p[j]-1),(a[i]+p[j]-1))))
   bet <- as.matrix( c(rep(0,a[i]+p[j]-1),(1:(a[i]+p[j]-1)) -
       mean(1:(a[i]+p[j]-1))) )
   bet <- bet/sum(bet^2)
```

# Loop 3, over lambda (variance ratios)
```
for(k in 1:nl){
```

# For the given lambda, compute M, where the vector $My$ contains the estimates of the intercept,
# age, period, and cohort effects in that order.
```
M <- solve.qr(qr(QpQ + lam[k]*D)) %*% t(Q)
```

# Compute alf'M and bet'M;  for each, save the maximum of the absolute values. These are
# hypothesized to be zero, and are, which implies the intercept and linear components are zero.
```
ndx <- ((i-1)*np + j-1)*nl + k
apmax[ ndx,c(1,2,3)] <- c(a[i],p[j],lam[k])
apmax[ndx,4] <- max(abs( t(alf) %*% M ))
apmax[ndx,5] <- max(abs( t(bet) %*% M ))
```

# end Loop 3
```
}
```
# end Loop 2
```
}
```
# end Loop 1
```
}
```

**Appendix B: R code for computing the fraction for a 2-RE-APC model with fixed age and random period and cohort effects.**

The matrix with which we began this construction is given below; it is the design matrix for an APC model with all three effects modeled as random, so each effect has the identity parameterization. The first column is the fixed intercept; columns 12-17, 18-22, and 2-11 are the columns in the design matrix for the age ($a = 6$), period ($p = 5$), and cohort ($c = 10$) effects, respectively. Beginning with this, we construct design matrices for the 2-RE-APC model.

```
LLYdm <- read.csv(file="LLYdm", header=FALSE, colClasses="numeric")
Nt <- as.matrix(LLYdm[,1])
Xa <- LLYdm[,12:17]
Zp <- LLYdm[,18:22]
Zc <- LLYdm[,2:11]
```

```
# Construct orthogonal components of age, period, and cohort design matrices.
# Apply an orthogonal transformation to each effect to get design-matrix columns
# corresponding to the intercept (level), linear component, quadratic component, etc.
Ga <- matrix(poly(1:6,5),6,5)
Ga <- cbind(as.matrix(rep(1/sqrt(6),6)),Ga)
Gp <- matrix(poly(1:5,4),5,4)
Gp <- cbind(as.matrix(rep(1/sqrt(5),5)),Gp)
Gc <- matrix(poly(1:10,9),10,9)
Gc <- cbind(as.matrix(rep(1/sqrt(10),10)),Gc)

Xa <- as.matrix(Xa) %*% Ga
Zp <- as.matrix(Zp) %*% Gp
Zc <- as.matrix(Zc) %*% Gc
```

```
# Regress the cohort effect's quadratic design-matrix column (Zc[,3]) on the age and
# period effects, which have also been re-parameterized to linear, quadratic, cubic, etc.
# components.
cquad <- lm(Zc[,3] ~ Xa + Zp).
```

```
# The fitted coefficients are given below. For this APC design,
# the fitted coefficients for the intercept is non-zero -0.145;
# for the age effect, the only non-zero coefficient is for the quadratic component (Xa3);
# for the period effect, the only non-zero coefficient is for the quadratic component(Zp3).
cquad$coef
```

[The output from this command follows]
> (Intercept)
> -1.450647e-01
> Xa1        Xa2           Xa3           Xa4           Xa5           Xa6
> NA  -9.121651e-17  2.659080e-01  -1.674049e-17  6.582039e-17  5.209998e-17
> Zp1        Zp2           Zp3           Zp4           Zp5
> NA   2.214129e-17  1.628347e-01  -6.798700e-17  -9.064933e-17

```
# The intercept piece of the cohort quadratic effect is the estimated intercept multiplied
# by a column of 30 1's, which is then squared and summed; the age piece is the squared
# length of column 3 of Xa multiplied by its coefficient, 2.659080e-01; the period piece
```

# is the squared length of column 3 of Zp multiplied by its coefficient 1.628347e-01;
# the part "left in the cohort effect" is the residual sum of squares from the regression
# above. That is, the cohort quadratic component's original squared length, 2.469697, is
# divided up as:
# intercept 0.631313, age 0.3535353, pd 0.1590908, left in cohort 1.325758, or somewhat
# more than half (54%)

# When the truth is that there are no age or period effects, the squared length in the
# cohort's quadratic component is m^2 2.469697.
# In computing the restricted likelihood (RL):
# m^2 (0.631313 + 0.3535353) is taken up by the fixed effects (i.e., is not available to the
# RL). Thus R2 for the regression of the cohort quadratic component on the fixed age
# effect and intercept is R2 = (0.631313 + 0.3535353)/2.469697 = 0.399, which rounds to
# 0.40. m^2 0.1590908 lies in the column space of the period effect, so it could arise
# from either period or cohort; 0.1590908/2.469697 is 6%. m^2 1.325758 is in the part
# of cohort's column space that's orthogonal to the age and period column spaces, so this
# is attributed only to cohort and 1.325758 / 2.469697 is 54%.

The matrix in the file LLYdm

| Intercept | C | | | | | | | | | | A | | | | | | P | | | |
|---|---|---|---|---|---|---|---|---|---|---|---|---|---|---|---|---|---|---|---|---|
| 1 | 0 | 0 | 0 | 0 | 0 | 1 | 0 | 0 | 0 | 0 | 1 | 0 | 0 | 0 | 0 | 0 | 1 | 0 | 0 | 0 | 0 |
| 1 | 0 | 0 | 0 | 0 | 0 | 0 | 1 | 0 | 0 | 0 | 1 | 0 | 0 | 0 | 0 | 0 | 0 | 1 | 0 | 0 | 0 |
| 1 | 0 | 0 | 0 | 0 | 0 | 0 | 0 | 1 | 0 | 0 | 1 | 0 | 0 | 0 | 0 | 0 | 0 | 0 | 1 | 0 | 0 |
| 1 | 0 | 0 | 0 | 0 | 0 | 0 | 0 | 0 | 1 | 0 | 1 | 0 | 0 | 0 | 0 | 0 | 0 | 0 | 0 | 1 | 0 |
| 1 | 0 | 0 | 0 | 0 | 0 | 0 | 0 | 0 | 0 | 1 | 1 | 0 | 0 | 0 | 0 | 0 | 0 | 0 | 0 | 0 | 1 |
| 1 | 0 | 0 | 0 | 0 | 1 | 0 | 0 | 0 | 0 | 0 | 0 | 1 | 0 | 0 | 0 | 0 | 1 | 0 | 0 | 0 | 0 |
| 1 | 0 | 0 | 0 | 0 | 0 | 1 | 0 | 0 | 0 | 0 | 0 | 1 | 0 | 0 | 0 | 0 | 0 | 1 | 0 | 0 | 0 |
| 1 | 0 | 0 | 0 | 0 | 0 | 0 | 1 | 0 | 0 | 0 | 0 | 1 | 0 | 0 | 0 | 0 | 0 | 0 | 1 | 0 | 0 |
| 1 | 0 | 0 | 0 | 0 | 0 | 0 | 0 | 1 | 0 | 0 | 0 | 1 | 0 | 0 | 0 | 0 | 0 | 0 | 0 | 1 | 0 |
| 1 | 0 | 0 | 0 | 0 | 0 | 0 | 0 | 0 | 1 | 0 | 0 | 1 | 0 | 0 | 0 | 0 | 0 | 0 | 0 | 0 | 1 |
| 1 | 0 | 0 | 0 | 1 | 0 | 0 | 0 | 0 | 0 | 0 | 0 | 0 | 1 | 0 | 0 | 0 | 1 | 0 | 0 | 0 | 0 |
| 1 | 0 | 0 | 0 | 0 | 1 | 0 | 0 | 0 | 0 | 0 | 0 | 0 | 1 | 0 | 0 | 0 | 0 | 1 | 0 | 0 | 0 |
| 1 | 0 | 0 | 0 | 0 | 0 | 1 | 0 | 0 | 0 | 0 | 0 | 0 | 1 | 0 | 0 | 0 | 0 | 0 | 1 | 0 | 0 |
| 1 | 0 | 0 | 0 | 0 | 0 | 0 | 1 | 0 | 0 | 0 | 0 | 0 | 1 | 0 | 0 | 0 | 0 | 0 | 0 | 1 | 0 |
| 1 | 0 | 0 | 0 | 0 | 0 | 0 | 0 | 1 | 0 | 0 | 0 | 0 | 1 | 0 | 0 | 0 | 0 | 0 | 0 | 0 | 1 |
| 1 | 0 | 0 | 1 | 0 | 0 | 0 | 0 | 0 | 0 | 0 | 0 | 0 | 0 | 1 | 0 | 0 | 1 | 0 | 0 | 0 | 0 |
| 1 | 0 | 0 | 0 | 1 | 0 | 0 | 0 | 0 | 0 | 0 | 0 | 0 | 0 | 1 | 0 | 0 | 0 | 1 | 0 | 0 | 0 |
| 1 | 0 | 0 | 0 | 0 | 1 | 0 | 0 | 0 | 0 | 0 | 0 | 0 | 0 | 1 | 0 | 0 | 0 | 0 | 1 | 0 | 0 |
| 1 | 0 | 0 | 0 | 0 | 0 | 1 | 0 | 0 | 0 | 0 | 0 | 0 | 0 | 1 | 0 | 0 | 0 | 0 | 0 | 1 | 0 |
| 1 | 0 | 0 | 0 | 0 | 0 | 0 | 1 | 0 | 0 | 0 | 0 | 0 | 0 | 1 | 0 | 0 | 0 | 0 | 0 | 0 | 1 |
| 1 | 0 | 1 | 0 | 0 | 0 | 0 | 0 | 0 | 0 | 0 | 0 | 0 | 0 | 0 | 1 | 0 | 1 | 0 | 0 | 0 | 0 |
| 1 | 0 | 0 | 1 | 0 | 0 | 0 | 0 | 0 | 0 | 0 | 0 | 0 | 0 | 0 | 1 | 0 | 0 | 1 | 0 | 0 | 0 |
| 1 | 0 | 0 | 0 | 1 | 0 | 0 | 0 | 0 | 0 | 0 | 0 | 0 | 0 | 0 | 1 | 0 | 0 | 0 | 1 | 0 | 0 |
| 1 | 0 | 0 | 0 | 0 | 1 | 0 | 0 | 0 | 0 | 0 | 0 | 0 | 0 | 0 | 1 | 0 | 0 | 0 | 0 | 1 | 0 |
| 1 | 0 | 0 | 0 | 0 | 0 | 1 | 0 | 0 | 0 | 0 | 0 | 0 | 0 | 0 | 1 | 0 | 0 | 0 | 0 | 0 | 1 |
| 1 | 1 | 0 | 0 | 0 | 0 | 0 | 0 | 0 | 0 | 0 | 0 | 0 | 0 | 0 | 0 | 1 | 1 | 0 | 0 | 0 | 0 |

```
1    0 1 0 0 0 0 0 0 0 0 0 0 0 0 1 0 1 0 0 0
1    0 0 1 0 0 0 0 0 0 0 0 0 0 0 1 0 0 1 0 0
1    0 0 0 1 0 0 0 0 0 0 0 0 0 1 0 0 0 1 0
1    0 0 0 0 1 0 0 0 0 0 0 0 0 1 0 0 0 0 1
```

**Appendix C: R code for generating datasets in Table 5 for 2-RE-APC models with fixed age effect and random period and cohort effects.**

The code below produced Table 5 on the second author's Macintosh laptop, but the reader may not get identical results even using the same random-number seed we used, because random number generators sometimes differ between machines or operating systems. However, using this code, the reader will get the same results with repeated runs because this code specifies the seed for the random number generator.

The matrix with which we began this construction is given in Appendix B; it is the design matrix for an APC model with all three effects modeled as random, so each effect has the identity parameterization. The first column is the fixed intercept; columns 12-17, 18-22, and 2-11 are the columns in the design matrix for the age ($a = 6$), period ($p = 5$), and cohort ($c = 10$) effects, respectively. Beginning with this, we construct design matrices for the 2-RE-APC model.

```
LLYdm <- read.csv(file="LLYdm", header=FALSE, colClasses="numeric")

Nt <- as.matrix(LLYdm[,1])
Xa <- LLYdm[,12:17]
Zp <- LLYdm[,18:22]
Zc <- LLYdm[,2:11]
```

# Construct orthogonal components of age, period, and cohort design matrices.
# Apply an orthogonal transformation to each effect to get design-matrix columns
# corresponding to the intercept (level), linear component, quadratic component, etc.
```
Ga <- matrix(poly(1:6,5),6,5)
Ga <- cbind(as.matrix(rep(1/sqrt(6),6)),Ga)
Gp <- matrix(poly(1:5,4),5,4)
Gp <- cbind(as.matrix(rep(1/sqrt(5),5)),Gp)
Gc <- matrix(poly(1:10,9),10,9)
Gc <- cbind(as.matrix(rep(1/sqrt(10),10)),Gc)

Xa <- as.matrix(Xa) %*% Ga
Zp <- as.matrix(Zp) %*% Gp
Zc <- as.matrix(Zc) %*% Gc
```

# Construct factors for use in fitting models using lmer
```
age <- rep(1:6,rep(5,6))
pd <- rep(1:5,6)
Z <- data.matrix(LLYdm[,2:11])
chrt <- Z %*% matrix(1:10,10,1)

age <- as.factor(age)
contrasts(age) <- contr.sum(6)
pd <- as.factor(pd)
chrt <- as.factor(chrt)
```

# Run the simulation that produced Table 5.
```
set.seed(5) # This sets the seed in rnorm's random number generator so the
            # simulation can be re-run and give identical results.
m <- 0 + (0:20)*0.05
```

# Save estimates of standard deviations (sigmas): Column 1 is m, Column 2 is the SD of

```r
# the cohort random effect, Column 3 is the SD of the period random effect, and Column
# 4 is the error SD
sdests <- matrix(NA,100*length(m),4)
# Save linear component estimates;  Column 1 is m, Column 2 is the estimated linear
# component of the age effect, Column 3 is the estimated linear component of the period
# effect, and Column 4 is the estimated linear component of the cohort effect
linests <- matrix(NA,100*length(m),4)

# The outer loop is over m, the inner loop is over replicate datasets for given m
for(i in 1:length(m)){
for(j in 1:100){

# Simulate the j-th dataset for a given value of m
y <- Zc[,2] + m[i]*Zc[,3] + rnorm(30,sd=0.01)
# fit the random-effects models
fit1 <- lmer(y ~ age + (1|pd) + (1|chrt))

# Save the estimated SDs
sdests[(i-1)*100+j,1] <- m[i]
sdests[(i-1)*100+j,2] <- sqrt(c(summary(fit1)$varcor$chrt))
sdests[(i-1)*100+j,3] <- sqrt(c(summary(fit1)$varcor$pd))
sdests[(i-1)*100+j,4] <- sigma(fit1)

# Save estimated linear component;  first m, then the age, period, and cohort effects
linests[(i-1)*100+j,1] <- m[i]
# linear component of the age effect
fage <- c(summary(fit1)$coef[2:6,1],-sum(summary(fit1)$coef[2:6,1]))
x <- 1:6
linests[(i-1)*100+j,2] <- lm(fage ~ x)$coef[2]
# linear component of the period effect
xt <- 1:5
linests[(i-1)*100+j,3] <- lm(data.matrix(ranef(fit1)$pd) ~ xt)$coef[2]
# linear component of the cohort effect
xt <- 1:10
linests[(i-1)*100+j,4] <- lm(data.matrix(ranef(fit1)$chrt) ~ xt)$coef[2]
}
}

# Count how many datasets had the linear component of the period effect shrunk really
# small.
tmp <- linests[,3] < 1e-2
table(linests[,1],tmp)
```